\documentclass[10pt]{article} 

\usepackage{amsmath}
\usepackage{amssymb}
\usepackage{graphics}
\usepackage[latin1]{inputenc}

\newtheorem{thm}{Theorem}
\newtheorem{prop}{Proposition}
\newtheorem{coro}{Corollary}
\newtheorem{lemma}{Lemma}
\newtheorem{remark}{Remark}
\newtheorem{definition}{Definition}

\newcommand{\R}{\mathbb{R}}             
\newcommand{\N}{\mathbb{N}}             
\newcommand{\Z}{\mathbb{Z}}             
\newcommand{\C}{\mathbb{C}}             
\renewcommand{\H}{\mathcal{H}}          
\newcommand{\G}{\mathcal{G}}   
\newcommand{\D}{\mathcal{D}}            
\newcommand{\B}{\mathcal{B}}            
\newcommand{\M}{\mathcal{M}}            

\newcommand{\co}{C_{0}^{\infty}}                  
\newcommand{\cor}{C_{0}^{\infty}(\R)}             

\newcommand{\e}{\epsilon}
\newcommand{\eo}{\epsilon_{0}}
\newcommand{\half}{\frac{1}{2}}

\newcommand{\Go}{\Gamma^0}         
\newcommand{\Ga}{\Gamma^1}         
\newcommand{\Gb}{\Gamma^2}         
\newcommand{\Gc}{\Gamma^3}         
\newcommand{\Gd}{\Gamma^5}         

\newcommand{\kl}{\langle}
\newcommand{\er}{\rangle}

\newcommand{\x}{\kl x \er}
\renewcommand{\r}{\kl r \er}

\newcommand{\DR}{\mathbb{D}}               
\newcommand{\DS}{\mathbb{D}_{S^3}}         
\newcommand{\DO}{\mathbb{D}_{0}}           

\newcommand{\Ra}{\mathcal{R}}              
\newcommand{\A}{\mathcal{A}}               

\newcommand{\Section}[1]{\section{#1} \setcounter{equation}{0}}

\begin{document}
\title{Local energy decay of massive Dirac fields in the 5D Myers-Perry metric}
\author{Thierry Daud\'e $^{\,1}$and Niky Kamran$^{\,2}$\\[12pt]
 $^1$  \small D\'epartement de Math\'ematiques \\
\small Universit\'e de Cergy-Pontoise \\
\small 95302 Cergy-Pontoise, France  \\
\small thierry.daude@u-cergy.fr\\
$^2$ \small Department of Mathematics and Statistics\\
\small  McGill
University\\ \small Montreal, QC, H3A 2K6, Canada\\
\small nkamran@math.mcgill.ca\\ } 




\date{}






\maketitle

\begin{abstract}
We consider massive Dirac fields evolving in the exterior region of a 5-dimensional Myers-Perry black hole and study their propagation properties. Our main result states that the local energy of such fields decays in a weak sense at late times. We obtain this result in two steps: first, using the separability of the Dirac equation, we prove the absence of a pure point spectrum for the corresponding Dirac operator; second, using a new form of the equation adapted to the local rotations of the black hole, we show by a Mourre theory argument that the spectrum is absolutely continuous. This leads directly to our main result.  
\end{abstract}


\Section{Introduction}

Since the publication of the seminal papers \cite{AADD, ADD, RaSu1, RaSu2}, higher dimensional black holes have attracted considerable attention, in particular in the context of brane-world theories \cite{Ma, Ka1}. In these scenarios, the physical world is represented by a four-dimensional brane embedded in a higher-dimensional bulk space-time whose geometry can be approximately described by the classical solutions of vacuum Einstein equations. A fascinating prediction of brane-world theories with large extra dimensions is the possibility of mini black hole production in high energy colliders such as the Large Hadron Collider \cite{DL, GT}, raising in turn the possibility of direct observations, for instance of Hawking radiation, and (possibly) of probes of large extra dimensions \cite{Ka2}. 

The metrics describing isolated rotating black holes in higher-dimensions were first constructed by Myers and Perry \cite{MP} as the asymptotically flat generalizations of four-dimensional Kerr vacuum solutions. Asymptotically (anti-) de Sitter rotating higher-dimensional black holes were then discovered by Hawking \textit{et al} in the 5D case \cite{HHTR} and by Gibbons \textit{et al.} in the general case \cite{GLPP1, GLPP2}. We refer to \cite{ER} for a review of  higher-dimensional black hole geometries. With this paper, we start a research program on the propagation and scattering properties of fields evolving in this type of geometries, both in the bulk and on four-dimensional branes embedded in the bulk. We are ultimately interested in extending the results of papers like \cite{Ba4, Ba5, Me3, Ha2, BaM, Dy, SZ, DaN1, DaN2} to this setting, that is to say in studying notions such as the Hawking effect, the resonances (or quasi-normal modes) and/or inverse scattering problems, phenomena or mechanisms that should provide a way to put in evidence the existence and influence of the extra dimensions of space-time.  

In this paper, we begin this program by considering massive Dirac fields that propagate in the bulk of a 5D Myers-Perry black hole and prove local decay of the energy at late times. In other words, we prove that the probability of finding Dirac particles in any compact region located outside the event horizon tends to zero as time $t$ goes to infinity. Our result can be stated more precisely as follows. Recall first that in Boyer-Linquist like coordinates $(t,r,\theta, \varphi,\psi)$, a 5D Myers-Perry black hole can be represented by the manifold
\begin{equation}
  \M = \R_t \times (0,+\infty)_r \times (0,\frac{\pi}{2})_\theta \times (0,2\pi)_\varphi \times (0,2\pi)_\psi,
\end{equation}
equipped with the Lorentzian metric (having signature $(-1,1,1,1,1)$)
\begin{equation} \label{Metric}
  g = - dt^2 + \frac{\Sigma r^2}{\Delta} dr^2 + \Sigma d\theta^2 + (r^2+a^2) \sin^2\theta d\varphi^2 + (r^2+b^2) \cos^2\theta d\psi^2  + \frac{\mu}{\Sigma} \big( dt - a \sin^2\theta d\varphi - b \cos^2\theta d\psi\big)^2,
\end{equation}
where 
\begin{equation} \label{DeltaSigma}
\begin{split}  
  \Delta & = (r^2+a^2)(r^2+b^2) - \mu r^2, \\
  \Sigma & = r^2 + a^2\cos^2\theta + b^2 \sin^2\theta.
\end{split}
\end{equation}
Myers-Perry black holes are completely determined by three parameters: their mass $\frac{\mu}{2}$ and the two independent angular momenta per unit mass, $a,b$, measured from infinity. The metric possesses three Killing vectors $\partial_t, \partial_\varphi, \partial_\psi$ reflecting the time-translation invariance and bi-azimuthal symmetry of the space-time. Note here that since the rotation group $SO(4)$ possesses two independent Casimir invariants, a rotating black hole in five dimensions may have two distinct planes of rotation specified by appropriate azimuthal coordinates - here $(\varphi, \psi)$ -, rather than a single axis of rotation. 

We shall restrict our attention to the non-extreme case $\mu > a^2 + b^2 + 2|ab|$, for which the function $\Delta$ has two distinct positive roots 
\begin{equation}
  r_\pm = \half \Big( \mu - a^2 - b^2 \pm \sqrt{(\mu -a^2-b^2)^2 - 4a^2 b^2} \Big),
\end{equation}
and can be written in factorized form as 
\begin{equation} \label{Delta}
  \Delta = (r^2-r_-^2)(r^2-r_+^2).
\end{equation}  
The radii $r_-$ and $r_+$ are called Cauchy and event horizons respectively and correspond to "coordinate" singularities of the metric. We shall in this paper only consider the exterior region of the black hole, that is the region $\{r > r_+\}$.    

We list here some important properties of the exterior region of a $5$D Myers-Perry black hole. First, it is of Petrov type D, so that its Weyl curvature possesses a pair of real principal null vector fields \cite{Wu} 
\begin{equation} \label{PNG}
  V^\pm = \frac{(r^2 + a^2)(r^2+b^2)}{\Delta} \Big(\partial_t + \frac{a}{r^2+a^2} \partial_\varphi + \frac{b}{r^2+b^2} \partial_\psi \Big) \pm \partial_r,
\end{equation}
that generate a congruence of shearfree null geodesics called principal null geodesics. Because of the factor $\frac{1}{\Delta}$ in (\ref{PNG}) which blows up when $r \to r_+$, we see that the principal null geodesics will not reach the event horizon in finite time $t$. As a  consequence, when described using the Boyer-Lindquist coordinate system, the event horizon is perceived as an asymptotic region of space-time. The geometry there turns out to be of asymptotically hyperbolic type. Conversely, when $r \to +\infty$, the metric (\ref{Metric}) tends to the Minkowski metric written in oblate bi-spheroidal coordinates. The geometry of the region $\{r=+\infty\}$, which corresponds to spatial infinity, is thus asymptotically flat. Second, the exterior region of a 5D Myers-Perry black hole is globally hyperbolic, the hypersurface $\Sigma_0 = \{t=0\} = (r_+,+\infty)_r \times S^3$ being a Cauchy hypersurface. We are thus able to express the Dirac equation as an evolution equation on the spatial hypersurface $\Sigma_0$ which, according to the above discussion, can be viewed as a cylindrical manifold having two ends - the event horizon $\{r=r_+\}$ and spatial infinity $\{r=+\infty\}$ - with very different geometries. We generically write this equation in Hamiltonian form as 
$$
  i\partial_t \psi = \DR \psi.
$$
The Hamiltonian $\DR$ acts on the Hilbert space $\H = L^2((r_+,+\infty) \times S^3, \textup{dVol}_\Sigma; \C^4)$ with dVol$_\Sigma$ the volume measure induced by $g$ on the hypersurface $\Sigma_0$. The main result of this paper is
\begin{thm} \label{MainThm}
  The Dirac operator $\DR$ is a self-adjoint operator on $\H$ having purely absolutely continuous spectrum. Moreover, for all $\chi \in \co((r_+,+\infty))$ and all $\psi \in \H$, we have the local energy decay
\begin{equation}
  \lim_{t \to \pm \infty} \| \chi(r) e^{-it\DR} \psi \| = 0.
\end{equation}
\end{thm}

Hence, the energy - or more precisely, the probability of presence - of massive Dirac fields cannot remain trapped in any compact set outside the event horizon when $t$ goes to infinity: the essential part of the fields must escape either to infinity, or to the event horizon at late times. We thus obtain analogous results to those proved for the Kerr metric \cite{Da3, FKSY2, FKSY3, HaN} in this new setting. 

The main ideas used in the proof of our main theorem are the following. Firstly, the absence of pure point spectrum of the Dirac operator $\DR$ relies crucially on the fact that the massive Dirac equation in the 5D Myers-Perry metric can be separated into purely radial and angular systems of ODEs \cite{FK, Wu}. Using a decomposition of the angular differential operator on a well chosen Hilbert basis, the problem amounts thus to proving the non-existence of $L^2$ solutions of a system of ODEs in the radial variable $r$. This, in turn, is shown using the particular nature of the geometry - of asympotically hyperbolic type - at the event horizon. Note that the spectrum of $\DR$ being continuous, we already get a local energy decay (in a $L^2$ mean sense) as a simple application of the RAGE theorem \cite{RS}. 

Secondly, we show the absence of continuous singular spectrum by means of a Mourre theory argument, a technique already used in the Kerr setting in \cite{Da4, HaN}. As in these papers, the expression of the Dirac equation which is adapted to the separation of variables turns out to be inconvenient for the construction of a locally conjugate operator. Indeed, the Dirac operator in the above separated form cannot be written as a short-range perturbation (of the same order) of a spherically symmetric like Dirac operator $\DO$, a situation in which Mourre theory typically works, but only as a long-range perturbation of $\DO$. These long-range terms are mainly due to the distant effects of the rotations of the space-time. We thus use another form of the equation that "minimizes" the effects of the rotations. This is done by choosing a local Lorentz frame in the calculation of the equation adapted to locally non rotating observers (see for instance \cite{AF}). We then obtain an expression of the Dirac equation that has the convenient form we expected. The new expression turns out to be quite similar to the one obtained in the Kerr case studied in \cite{Da4} and we are able to construct a locally conjugate operator for $\DR$ and prove the absence of continuous singular spectrum. Finally, our local energy decay is then a simple consequence of the Riemann-Lebesgue lemma. 

This paper is organized as follows. In Section \ref{MP-BlackHoles}, we first briefly present the Cartan formalism to calculate the Dirac equation in a curved $5$D-space-time. We then find an expression of the Dirac equation in the Myers-Perry metric based on the choice of a local Lorentz frame adapted to the point of view of locally non-rotating observers. We conclude this section by various simplifications of the initial equation that is finally put in a convenient Hamiltonian form. In Section \ref{SpectralProperties}, we gather all the needed analytical and spectral properties of the Hamiltonian $\DR$ needed in the later Mourre analysis. We also use here the absence of pure point spectrum for $\DR$ (whose proof is given in Appendix \ref{Separation}) to obtain our first local energy decay in an $L^2$ mean sense. Finally, in Section \ref{Mourre}, we first briefly recall the basics of Mourre theory and then construct a locally conjugate operator for the Hamiltonian $\DR$. We conclude the paper by proving our main Theorem \ref{MainThm}.


\Section{The Dirac equation in the $5$D-Myers-Perry metric} \label{MP-BlackHoles}

\subsection{Orthonormal frame formalism for the Dirac equation in $5$D curved space-time}
  
To calculate the massive Dirac equation in a 5D Myers-Perry black hole, we use Cartan's orthonormal frame formalism. Let us denote by $\{e_A\}_{A=0,1,2,3,5}$ a given local Lorentz frame, \textit{i.e.} satisfying $g(e_A, e_B) = \eta_{AB}$ where $\eta_{AB} =$ diag$(-1,1,1,1,1)$ is the flat (Lorentz) metric. We also denote by $\{e^A\}_{A=0,1,2,3,5}$ the set of dual $1$-forms of the frame $\{e_A\}$. Latin letters A,B will denote in what follows local Lorentz frame indices, while Greek letters $\mu, \nu$ run over five-dimensional space-time coordinates indices $(t,r,\theta, \varphi, \psi)$. The \emph{massive} Dirac equation takes then the generic form
\begin{equation} \label{AbstractDiracEq}
  (\mathbb{H} + m) \phi = [\gamma^A e_A^\mu (\partial_\mu + \Gamma_\mu) + m] \phi = 0. 
\end{equation}  
Here, the $\gamma^A$'s are the gamma Dirac matrices satisfying the anticommutation relations 
\begin{equation} \label{Clifford}
  \{\gamma^A, \gamma^B\} = \gamma^A \gamma^B + \gamma^B \gamma^A = 2 \eta^{AB}, 
\end{equation} 
and $\Gamma = \Gamma_\mu dx^\mu = \Gamma_A e^A$ stands for the spinor connection $1$-form. In order to derive it, we first compute the spin-connection $1$-form $\omega_{AB} = \omega_{AB\mu} dx^\mu = f_{ABC} e^C$ thanks to Cartan's first structural equation
\begin{equation} \label{Cartan}
  d e^A + \omega^A_{\ B} \wedge e^B = 0, \quad \omega_{AB} = \eta_{AC} \omega^C_{\ B} = - \omega_{BA}. 
\end{equation} 
To obtain the spinor connection $1$-form $\Gamma$ from $\omega_{AB}$, we make use of the homomorphism between the $SO(4,1)$ group and its spinor representation which is derived from the relation (\ref{Clifford}). The $so(4,1)$ Lie algebra is defined by the ten antisymmetric generators $\Sigma^{AB} = [\gamma^A,\gamma^B]/(2i)$ which gives the spinor representation, and the spinor connection can be viewed as a $so(4,1)$ Lie-algebra-valued $1$-form. Using the isomorphism between the $so(4,1)$ Lie algebra and its spinor representation, \textit{i.e.} $\Gamma_\mu = (i/4) \Sigma^{AB} \omega_{AM\mu} = (1/4) \gamma^A \gamma^B \omega_{AB\mu}$, we can construct the spinor connection $1$-form by
\begin{equation}
  \Gamma = \frac{1}{8} [\gamma^A, \gamma^B] \omega_{AB} = \frac{1}{4} \gamma^A \gamma^B \omega_{AB} = \frac{1}{4} \gamma^A \gamma^B f_{ABC} e^C. 
\end{equation}   
Now in terms of the local differential operators $\partial_A = e_A^\mu \partial_\mu$, the Dirac equation (\ref{AbstractDiracEq}) can be rewritten in the local Lorentz frame as
\begin{equation} \label{DE}
  [\gamma^A ( \partial_A + \Gamma_A) + m] \phi = 0,
\end{equation}
where the $\Gamma_A = e_A^\mu \Gamma_\mu = \frac{1}{4} \gamma^B \gamma^C f_{BCA}$ are the components of the spinor connection in the local Lorentz frame. Finally, note that the Clifford algebra in dimension five has two different reducible representations (they differ by the factor of a $\gamma^5$ matrix). It is usually assumed that fermion fields are in a reducible representation of the Clifford algebra. In other words, we can work with the Dirac equation in a four-component spinor formalism like in the four-dimensional case, and just need to take the $\gamma^5$ matrix as the fifth basis vector component.

\subsection{The Dirac equation adapted to locally non rotating observers}

To calculate the expression of the Dirac operator, we first need to choose a local Lorentz frame $\{e_A\}_{A=0,1,2,3,5}$. Note that to any local Lorentz frame there corresponds a particular local observer given by the timelike unit vector field $e_0$. When dealing with the Dirac equation in rotating black holes, it is common to use the Petrov type $D$ character of the space-time and the existence of the principal null geodesics generated by (\ref{PNG}) to define a local observer. Recall that in the case of $5$D-Myers-Perry black holes, the principal null geodesics are generated by the pair of real principal null vectors 
$$
  V^\pm = \frac{(r^2 + a^2)(r^2+b^2)}{\Delta} \Big(\partial_t + \frac{a}{r^2+a^2} \partial_\varphi + \frac{b}{r^2+b^2} \partial_\psi \Big) \pm \partial_r,
$$
whose normalized sum defines a local observer 
$$
  U = \frac{(r^2 + a^2)(r^2+b^2)}{\Delta} \Big(\partial_t + \frac{a}{r^2+a^2} \partial_\varphi + \frac{b}{r^2+b^2} \partial_\psi \Big).
$$ 
The interest in choosing a local Lorentz frame adapted to the observers described by $U$ is that the corresponding expression of the massive Dirac equation allows for separation of variables. In Appendix \ref{Separation}, we recall this expression obtained by Wu in \cite{Wu} and use it to prove the absence of eigenmodes. 

However, as mentioned in the Introduction, this form of the Dirac equation turns out not to be convenient for our purpose since the resulting Hamiltonian cannot be written as a short-range perturbation of a spherically symmetric like Dirac operator. Following \cite{Da3, HaN}, we thus choose another local Lorentz frame based on locally non rotating observers. These are described by the vector field $T$ orthogonal to the spatial hypersurfaces $\Sigma_t = \{t= \textrm{const}\}$ and normalized such that $g(T,T) = -1$. To compute the vector field $T$, we introduce some notation. The coefficients of the metric (\ref{Metric}) in the coordinate basis are given by
\begin{equation}
\begin{split}
  g_{tt} & = -1 + \frac{\mu}{\Sigma}, \quad g_{rr} = \frac{r^2\Sigma}{\Delta}, \quad g_{\theta \theta} = \Sigma, \quad g_{\varphi \varphi} = \frac{\alpha \sin^2\theta}{\Sigma}, \quad g_{\psi \psi} = \frac{\beta \cos^2\theta}{\Sigma}, \\
  g_{t\varphi} & = - \frac{2a\mu\sin^2\theta}{\Sigma}, \quad  g_{t\psi} = - \frac{2b\mu\cos^2\theta}{\Sigma}, \quad g_{\varphi \psi} = \frac{2ab\mu\sin^2\theta \cos^2\theta}{\Sigma},  
\end{split}  
\end{equation}      
where
\begin{equation}
 \alpha = (r^2+a^2) \Sigma + a^2 \mu \sin^2 \theta, \quad \beta = (r^2+b^2) \Sigma + b^2 \mu \cos^2 \theta.
\end{equation}  
The coefficients of the inverse metric $g^{-1}$ are found to be
\begin{equation}
\begin{split}
  g^{tt} & = - \frac{\sigma}{\Delta\Sigma}, \quad g^{rr} = \frac{\Delta}{r^2\Sigma}, \quad g^{\theta \theta} = \frac{1}{\Sigma}, \\
  g^{\varphi \varphi} & = \frac{1}{\tau} \Big( \frac{1}{\sin^2\theta} + \frac{(r^2+b^2)(b^2-a^2) - \mu b^2}{\Delta} \Big), \quad
  g^{\psi \psi} = \frac{1}{\Sigma} \Big( \frac{1}{\cos^2\theta} + \frac{(r^2+a^2)(a^2-b^2) - \mu a^2}{\Delta} \Big), \\    
  g^{t\varphi} & = - \frac{\mu a (r^2+b^2)}{\Delta\Sigma}, \quad  g^{t\psi} = - \frac{\mu b (r^2+a^2)}{\Delta \Sigma}, \quad g^{\varphi \psi} = - \frac{\mu ab}{\Delta \Sigma},  
\end{split}  
\end{equation} 
where $\tau = \Delta \Sigma + \mu (r^2+a^2)(r^2+b^2)$. 

\begin{remark}
  For later reference, we list several identities that will be systematically used in the course of the calculations,
  \begin{equation}
  \begin{split}
    \tau & = \Delta \Sigma + \mu (r^2+a^2)(r^2+b^2) \\
         & = (r^2+a^2)(r^2+b^2)\Sigma + (r^2+a^2)\beta + (r^2+b^2)\alpha, \\
         & = (r^2+a^2)(r^2+b^2)\Sigma + a^2 \mu \sin^2\theta (r^2+b^2) + b^2 \mu \cos^2\theta (r^2+a^2), \\
    \Sigma \tau & = \alpha \beta - a^2 b^2 \mu^2 \cos^2\theta \sin^2\theta.
  \end{split} 
  \end{equation}
\end{remark}
  
The vector field $T$ orthogonal to the $\Sigma_t$ is collinear to $\nabla t$. In the basis associated to the Boyer-Lindquist coordinate system, we thus have 
$$
 \nabla^{\mu} t = g^{\mu \nu} \nabla_\nu t = g^{\mu \nu} (dt)_\nu = g^{t\mu}.
$$
Taking into account the normalization condition $g(T,T) = -1$, we find 
\begin{equation}
  T = \sqrt{\frac{\tau}{\Delta \Sigma}} \Big( \partial_t + \Omega_a \partial_\varphi + \Omega_b \partial_\psi \Big),
\end{equation}      
where
\begin{equation}
  \Omega_a = \frac{\mu a (r^2+b^2)}{\tau}, \quad \Omega_b = \frac{\mu b (r^2+a^2)}{\tau}. 
\end{equation}
     
\begin{remark}     
  Note that the functions $\Omega_a, \Omega_b$ tend to the constant values $\omega_a = a/(r_+^2 + a^2)$ and $\omega_b =    b/(r_+^2 + b^2)$ when $r$ tends to $r_+$ (the event horizon) whereas $\Omega_a, \Omega_b$ tend to $0$ when $r \to           \infty$. This illustrates a remarkable property of Myers-Perry black hole, namely the dragging of inertial frames in both the $\varphi$ and $\psi$ $2$-planes of rotation. The non-zero quantities $\omega_a, \omega_b$ can thus be interpreted as the angular velocities of the black hole horizon.      
\end{remark}     

We now choose the following local Lorentz frame corresponding to locally non rotating observers moving in the $\varphi$ $2$-plane \cite{AF}:
\begin{equation} \label{LLF}
\begin{split}
  e_0 & = T = \sqrt{\frac{\tau}{\Delta \Sigma}} \Big( \partial_t + \Omega_a ,\partial_\varphi + \Omega_b \,\partial_\psi \Big), \quad e_1  = \sqrt{\frac{\Delta}{\Sigma r^2}} \,\partial_r, \quad e_2  = \frac{1}{\sqrt{\Sigma}} \,\partial_\theta, \\
  e_3  & = \sqrt{\frac{\beta}{\tau \sin^2\theta}} \Big(\partial_\varphi - \frac{ab\mu \sin^2\theta}{\beta} \partial_\psi \Big), \quad e_5  = \sqrt{\frac{\Sigma}{\beta \cos^2\theta}} \,\partial_\psi. 
\end{split}
\end{equation}
The basis of dual $1$-forms is given by
\begin{equation}
\begin{split}
  e^0 & = \sqrt{\frac{\Delta \Sigma}{\tau}} \,dt, \quad e^1 = \sqrt{\frac{\Sigma r^2}{\Delta}} \,dr, \quad e^2 = \sqrt{\Sigma} \,d\theta, \quad e^3 = \sqrt{\frac{\tau \sin^2\theta}{\beta}} \Big(d\varphi - \Omega_a \,dt \Big), \\
  e^5 & = \sqrt{\frac{\beta \cos^2\theta}{\Sigma}} \Big(d\psi - \frac{b\mu}{\beta} \,dt + \frac{ab\mu \sin^2\theta}{\beta} \, d\varphi \Big). 
\end{split}
\end{equation}
After considerable algebraic manipulations, the exterior differential of the $1$-forms $\{e^A\}$ can be expressed as
\begin{equation}
\begin{split}
 d e^0 & = - \frac{\sqrt{\Delta}}{2 r \sqrt{\Sigma}} \Big( \frac{\partial_r \Delta}{\Delta} + \frac{\partial_r \Sigma}{\Sigma} - \frac{\partial_r \tau}{\tau} \Big) \, e^0 \wedge e^1 - \frac{1}{2 \sqrt{\Sigma}} \Big( \frac{\partial_\theta \Sigma}{\Sigma} - \frac{\partial_\theta \tau}{\tau} \Big) \, e^0 \wedge e^2, \\
  d e^1 & = - \frac{(\partial_\theta \Sigma)}{2 \Sigma^{\frac{3}{2}}} \ e^1 \wedge e^2, \\
  d e^2 & = \frac{\sqrt{\Delta}}{\Sigma^{\frac{3}{2}}} \ e^1 \wedge e^2, \\
  d e^3 & = \frac{\tau \sin \theta}{r \Sigma \sqrt{\beta}} (\partial_r \Omega_a) \, e^0 \wedge e^1 + \frac{\tau \sin \theta}{\Sigma \sqrt{\Delta \beta}} (\partial_\theta \Omega_a) \, e^0 \wedge e^2, \\
        & \ + \half \sqrt{\frac{\Delta}{\Sigma r^2}} \Big( \frac{\partial_r \tau}{\tau} - \frac{\partial_r \beta}{\beta} \Big) \, e^1 \wedge e^3 + \frac{1}{2 \sqrt{\Sigma}} \Big( \frac{\partial_\theta \tau}{\tau} + 2\cot \theta - \frac{\partial_\theta \beta}{\beta} \Big) \, e^2 \wedge e^3 \\
  d e^5 & = -\frac{b\mu(r^2+a^2) (\partial_r \beta) \cos\theta}{r \Sigma^{\frac{3}{2}} \sqrt{\beta \tau}} \, e^0\wedge e^1 - \frac{2b\mu \sin\theta \cos^2\theta \sqrt{\Delta} (b^2 - a^2)}{\Sigma^{\frac{3}{2}} \sqrt{\beta \tau}} \, e^0\wedge e^2 \\
        & \quad - \frac{ab\mu \sqrt{\Delta} (\partial_r \beta) \sin\theta \cos \theta}{\Sigma \beta r \sqrt{\tau}} \, e^1\wedge e^3 + \frac{\sqrt{\Delta}}{2r\sqrt{\Sigma}} \Big( \frac{(\partial_r \beta)}{\beta} - \frac{(\partial_r \Sigma)}{\Sigma}\Big) \, e^1\wedge e^5 \\
        & \quad + \frac{a b \mu \cos\theta \sin\theta}{\Sigma \sqrt{\tau}} \Big( 2\cot\theta - \frac{(\partial_\theta \beta)}{\beta} \Big) \, e^2\wedge e^3 + \frac{1}{2\sqrt{\Sigma}} \Big( \frac{(\partial_\theta \beta)}{\beta} - 2\tan\theta - \frac{(\partial_\theta \sigma)}{\sigma} \Big) \, e^2 \wedge e^5.  
\end{split}
\end{equation}
The spin-connection $1$-forms $\omega^A_{\ B}$ are now found from the Cartan's first structural equation (\ref{Cartan}) as follows

\begin{equation}
\begin{split}
  \omega^0_{\ 1} & =  \frac{\sqrt{\Delta}}{2 r \sqrt{\Sigma}} \Big( \frac{\partial_r \Delta}{\Delta} + \frac{\partial_r \Sigma}{\Sigma} - \frac{\partial_r \tau}{\tau} \Big) \, e^0 + \frac{\tau \sin \theta}{2 r \Sigma \sqrt{\beta}} (\partial_r \Omega_a) \, e^3 -\frac{b\mu(r^2+a^2) (\partial_r \beta) \cos\theta}{2 r \Sigma^{\frac{3}{2}} \sqrt{\beta \tau}} \, e^5, \\ 
  \omega^0_{\ 2} & =  \frac{1}{2 \sqrt{\Sigma}} \Big( \frac{\partial_\theta \Sigma}{\Sigma} - \frac{\partial_\theta \tau}{\tau} \Big) \, e^0 + \frac{\tau \sin \theta}{2 \Sigma \sqrt{\Delta \beta}} (\partial_\theta \Omega_a) \, e^3 - \frac{b\mu \sin\theta \cos^2\theta \sqrt{\Delta} (b^2 - a^2)}{\Sigma^{\frac{3}{2}} \sqrt{\beta \tau}} \, e^5, \\ 
  \omega^0_{\ 3} & =  \frac{\tau \sin \theta}{2 r \Sigma \sqrt{\beta}} (\partial_r \Omega_a) \, e^1 + \frac{\tau \sin \theta}{2 \Sigma \sqrt{\Delta \beta}} (\partial_\theta \Omega_a) \, e^2, \\
  \omega^0_{\ 5} & =  -\frac{b\mu(r^2+a^2) (\partial_r \beta) \cos\theta}{2 r \Sigma^{\frac{3}{2}} \sqrt{\beta \tau}} \, e^1 + \frac{b\mu \sin\theta \cos^2\theta \sqrt{\Delta} (b^2 - a^2)}{\Sigma^{\frac{3}{2}} \sqrt{\beta \tau}} \, e^2, \\
  \omega^1_{\ 2} & =  \frac{(\partial_\theta \Sigma)}{2 \Sigma^{\frac{3}{2}}} \ e^1 - \frac{\sqrt{\Delta}}{\Sigma^{\frac{3}{2}}} \ e^2, \\
  \omega^1_{\ 3} & =  \frac{\tau \sin \theta}{2 r \Sigma \sqrt{\beta}} (\partial_r \Omega_a) \, e^0 - \half \sqrt{\frac{\Delta}{\Sigma r^2}} \Big( \frac{\partial_r \tau}{\tau} - \frac{\partial_r \beta}{\beta} \Big) \, e^3 + \frac{ab\mu \sqrt{\Delta} (\partial_r \beta) \sin\theta \cos \theta}{2 \Sigma \beta r \sqrt{\tau}} \, e^5, \\
  \omega^1_{\ 5} & = -\frac{b\mu(r^2+a^2) (\partial_r \beta) \cos\theta}{2 r \Sigma^{\frac{3}{2}} \sqrt{\beta \tau}} \, e^0 + \frac{ab\mu \sqrt{\Delta} (\partial_r \beta) \sin\theta \cos \theta}{2 \Sigma \beta r \sqrt{\tau}} \, e^3 - \frac{\sqrt{\Delta}}{2r\sqrt{\Sigma}} \Big( \frac{(\partial_r \beta)}{\beta} - \frac{(\partial_r \Sigma)}{\Sigma}\Big) \, e^5, \\ 
  \omega^2_{\ 3} & =  \frac{\tau \sin \theta}{2 \Sigma \sqrt{\Delta \beta}} (\partial_\theta \Omega_a) \, e^0 + \frac{1}{2 \sqrt{\Sigma}} \Big( \frac{\partial_\theta \tau}{\tau} + 2\cot \theta - \frac{\partial_\theta \beta}{\beta} \Big) \, e^3 - \frac{a b \mu \cos\theta \sin\theta}{2 \Sigma \sqrt{\tau}} \Big( 2\cot\theta - \frac{(\partial_\theta \beta)}{\beta} \Big) \, e^5, \\
  \omega^2_{\ 5} & =  - \frac{b\mu \sin\theta \cos^2\theta \sqrt{\Delta} (b^2 - a^2)}{\Sigma^{\frac{3}{2}} \sqrt{\beta \tau}} \, e^0 - \frac{a b \mu \cos\theta \sin\theta}{2 \Sigma \sqrt{\tau}} \Big( 2\cot\theta - \frac{(\partial_\theta \beta)}{\beta} \Big) \, e^3 \\
                 &  \hspace{7cm} - \frac{1}{2\sqrt{\Sigma}} \Big( \frac{(\partial_\theta \beta)}{\beta} - 2\tan\theta - \frac{(\partial_\theta \sigma)}{\sigma} \Big) \, e^5, \\
  \omega^3_{\ 5} & = - \frac{ab\mu \sqrt{\Delta} (\partial_r \beta) \sin\theta \cos \theta}{2 \Sigma \beta r \sqrt{\tau}} \, e^1 + \frac{a b \mu \cos\theta \sin\theta}{2 \Sigma \sqrt{\tau}} \Big( 2\cot\theta - \frac{(\partial_\theta \beta)}{\beta} \Big) \, e^2.
\end{split} 
\end{equation}
We then deduce the local Lorentz frame component $\Gamma_A$ from the spinor connection $1$-form 
$$
 \Gamma = \Gamma_A e^A = (1/4) \gamma^A \gamma^B \omega_{AB}.
$$ 
We get

\begin{equation} \label{SpinorConnection}
\begin{split}
  \Gamma_0 & = \half \Big[ - \frac{\sqrt{\Delta}}{2 r \sqrt{\Sigma}} \Big( \frac{\partial_r \Delta}{\Delta} + \frac{\partial_r \Sigma}{\Sigma} - \frac{\partial_r \tau}{\tau} \Big) \gamma^0 \gamma^1 - \frac{1}{2 \sqrt{\Sigma}} \Big( \frac{\partial_\theta \Sigma}{\Sigma} - \frac{\partial_\theta \tau}{\tau} \Big)  \gamma^0 \gamma^2 + \frac{\tau \sin \theta}{2 r \Sigma \sqrt{\beta}} (\partial_r \Omega_a) \gamma^1 \gamma^3  \\
           & \hspace{1cm} -\frac{b\mu(r^2+a^2) (\partial_r \beta) \cos\theta}{2 r \Sigma^{\frac{3}{2}} \sqrt{\beta \tau}} \gamma^1 \gamma^5 + \frac{\tau \sin \theta}{2 \Sigma \sqrt{\Delta \beta}} (\partial_\theta \Omega_a) \gamma^2 \gamma^3  - \frac{b\mu \sin\theta \cos^2\theta \sqrt{\Delta} (b^2 - a^2)}{\Sigma^{\frac{3}{2}} \sqrt{\beta \tau}} \gamma^2 \gamma^5  \Big], \\
\Gamma_1 & = \half \Big[ \gamma^0 \gamma^3 + \frac{b\mu(r^2+a^2) (\partial_r \beta) \cos\theta}{2 r \Sigma^{\frac{3}{2}} \sqrt{\beta \tau}}  \gamma^0 \gamma^5 + \frac{(\partial_\theta \Sigma)}{2 \Sigma^{\frac{3}{2}}} \gamma^1 \gamma^2 - \frac{ab\mu \sqrt{\Delta} (\partial_r \beta) \sin\theta \cos \theta}{2 \Sigma \beta r \sqrt{\tau}} \gamma^3 \gamma^5 \Big], \\
\Gamma_2 & = \half \Big[ - \frac{\tau \sin \theta}{2 \Sigma \sqrt{\Delta \beta}} (\partial_\theta \Omega_a) \gamma^0 \gamma^3 + \frac{b\mu \sin\theta \cos^2\theta \sqrt{\Delta} (b^2 - a^2)}{\Sigma^{\frac{3}{2}} \sqrt{\beta \tau}}  \gamma^0 \gamma^5 - \frac{\sqrt{\Delta}}{\Sigma^{\frac{3}{2}}} \gamma^1 \gamma^2 \\
           & \hspace{3cm} + \frac{a b \mu \cos\theta \sin\theta}{2 \Sigma \sqrt{\tau}} \Big( 2\cot\theta - \frac{(\partial_\theta \beta)}{\beta} \Big) \gamma^3 \gamma^5 \Big], \\
\Gamma_3 & = \half \Big[ - \frac{\tau \sin \theta}{2 r \Sigma \sqrt{\beta}} (\partial_r \Omega_a) \gamma^0 \gamma^1 - \frac{\tau \sin \theta}{2 \Sigma \sqrt{\Delta \beta}} (\partial_\theta \Omega_a) \gamma^0 \gamma^2 - \half \sqrt{\frac{\Delta}{\Sigma r^2}} \Big( \frac{\partial_r \tau}{\tau} - \frac{\partial_r \beta}{\beta} \Big)  \gamma^1 \gamma^3 \\
         & \hspace{3cm} + \frac{ab\mu \sqrt{\Delta} (\partial_r \beta) \sin\theta \cos \theta}{2 \Sigma \beta r \sqrt{\tau}}   \gamma^1 \gamma^5 - \frac{1}{2 \sqrt{\Sigma}} \Big( \frac{\partial_\theta \tau}{\tau} + 2\cot \theta - \frac{\partial_\theta \beta}{\beta} \Big) \gamma^2 \gamma^3 \\
         & \hspace{5cm} - \frac{a b \mu \cos\theta \sin\theta}{2 \Sigma \sqrt{\tau}} \Big( 2\cot\theta - \frac{(\partial_\theta \beta)}{\beta} \Big)  \gamma^2 \gamma^5  \Big], \\
\Gamma_5 & = \half \Big[ \frac{b\mu(r^2+a^2) (\partial_r \beta) \cos\theta}{2 r \Sigma^{\frac{3}{2}} \sqrt{\beta \tau}}\gamma^0 \gamma^1 + \frac{b\mu \sin\theta \cos^2\theta \sqrt{\Delta} (b^2 - a^2)}{\Sigma^{\frac{3}{2}} \sqrt{\beta \tau}} \gamma^0 \gamma^2 \\
         & \hspace{1.5cm} + \frac{ab\mu \sqrt{\Delta} (\partial_r \beta) \sin\theta \cos \theta}{2\Sigma \beta r \sqrt{\tau}} \gamma^1 \gamma^3 - \frac{\sqrt{\Delta}}{2r\sqrt{\Sigma}} \Big( \frac{(\partial_r \beta)}{\beta} - \frac{(\partial_r \Sigma)}{\Sigma}\Big) \gamma^1 \gamma^5 \\
         & \hspace{2cm} - \frac{a b \mu \cos\theta \sin\theta}{2 \Sigma \sqrt{\tau}} \Big( 2\cot\theta - \frac{(\partial_\theta \beta)}{\beta} \Big) \gamma^2 \gamma^3 - \frac{1}{2\sqrt{\Sigma}} \Big( \frac{(\partial_\theta \beta)}{\beta} - 2\tan\theta - \frac{(\partial_\theta \sigma)}{\sigma} \Big) \gamma^2 \gamma^5  \Big].   
\end{split}
\end{equation}
Moreover, the differential part of the Dirac operator (\ref{DE}) is 

\begin{equation}
\begin{split}
  \gamma^A \partial_A & = \gamma^0 \sqrt{\frac{\tau}{\Delta \Sigma}} \Big( \partial_t + \Omega_a \partial_\varphi + \Omega_b \partial_\psi \Big) + \gamma^2 \sqrt{\frac{\Delta}{r^2 \Sigma}} \partial_r + \gamma^2 \frac{1}{\sqrt{\Sigma}} \partial_\theta \\
                      & \hspace{2cm} + \gamma^3 \sqrt{\frac{\beta}{\tau \sin^2\theta}} \Big(\partial_\varphi - \frac{ab\mu\sin^2\theta}{\beta} \partial_\psi \Big) + \gamma^5 \sqrt{\frac{\Sigma}{\beta \cos^2\theta}} \partial_\psi. 
\end{split}
\end{equation}
We are now able to give the expression of the full Dirac equation (\ref{DE}) in the particular local Lorentz frame (\ref{LLF}). We have  

\begin{equation}
\begin{split}
  \Big[ \gamma^0 \sqrt{\frac{\tau}{\Delta \Sigma}} \Big( \partial_t & + \Omega_a \partial_\varphi + \Omega_b \partial_\psi \Big) + \gamma^2 \sqrt{\frac{\Delta}{r^2 \Sigma}} \partial_r + \gamma^2 \frac{1}{\sqrt{\Sigma}} \partial_\theta \\
            & \hspace{1cm} + \gamma^3 \sqrt{\frac{\beta}{\tau \sin^2\theta}} \Big(\partial_\varphi - \frac{ab\mu\sin^2\theta}{\beta} \partial_\psi \Big) + \gamma^5 \sqrt{\frac{\Sigma}{\beta \cos^2\theta}} \partial_\psi + \gamma^A \Gamma_A + m \Big] \phi = 0. 
\end{split}
\end{equation}
Multiplying on the left by $-i\gamma^0 \sqrt{\frac{\tau}{\Delta \Sigma}}$, we obtain

\begin{equation}
\begin{split}
  i \partial_t \phi & = \Big[ i\gamma^0 \gamma^1 \frac{\Delta}{r \sqrt{\tau}} \partial_r + i \gamma^0 \gamma^2 \sqrt{\frac{\delta}{\tau}} \partial_\theta - i\Big( \Omega_a - \gamma^0 \gamma^3 \frac{\sqrt{\Delta \Sigma \beta}}{\tau \sin\theta} \Big) \partial_\varphi \\
                    & \quad - i\Big( \Omega_b + \gamma^0 \gamma^3 \frac{ab\mu\sqrt{\Delta \Sigma}\sin\theta}{\tau \sqrt{\beta}} - \gamma^0 \gamma^5 \frac{\sqrt{\Delta}\Sigma}{\sqrt{\beta \tau}\cos\theta} \Big) \partial_\psi + i \gamma^0 \sqrt{\frac{\Delta \Sigma}{\tau}} (\gamma^A \Gamma_A) + i m \gamma^0 \sqrt{\frac{\Delta \Sigma}{\tau}} \Big] \phi
\end{split}                    
\end{equation}
Using (\ref{SpinorConnection}) and the fact that $\gamma^5 = -i\gamma^0 \gamma^1 \gamma^2 \gamma^3$, we get for the connection term 

\begin{equation} 
\begin{split}
  i \gamma^0 \sqrt{\frac{\Delta \Sigma}{\tau}} (\gamma^A \Gamma_A) & = i \frac{\Delta}{4r\sqrt{\tau}} \Big( \frac{\partial_r \Delta}{\Delta} + \frac{\partial_r \Sigma}{\Sigma} \Big) \gamma^0 \gamma^1 + i \frac{\sqrt{\Delta}}{4\sqrt{\tau}} \Big( \frac{\partial_\theta \Sigma}{\Sigma} + 2 \cot\theta - 2 \tan\theta \Big) \gamma^0 \gamma^2 \\
    & \hspace{0.5cm} - i \frac{\sqrt{\Delta \tau} \sin\theta (\partial_r\Omega_a)}{4r\sqrt{\beta\Sigma}} \gamma^1 \gamma^3 - i \frac{\sqrt{\tau} \sin\theta (\partial_\theta\Omega_a)}{4\sqrt{\beta\Sigma}} \gamma^2 \gamma^3 \\
    & \hspace{1cm} + i \frac{b\mu(r^2+a^2) \sqrt{\Delta}\cos\theta(\partial_r\beta)}{4r\Sigma \tau \sqrt{\beta}} \gamma^1\gamma^5 + i \frac{b\mu\Delta(b^2-a^2)\cos^2\theta\sin\theta}{2\Sigma\tau\sqrt{\beta}} \gamma^2\gamma^5 \\
            & \hspace{1.5cm} + \frac{ab\mu\Delta(\partial_r\beta) \sin\theta\cos\theta}{4r\beta\tau\sqrt{\Sigma}} \gamma^2 + \frac{ab\mu\sqrt{\Delta} \sin\theta\cos\theta}{4\tau\sqrt{\Sigma}} \gamma^1,  
\end{split}
\end{equation}
an expression that we abbreviate as
\begin{equation} \label{V0}
  i \gamma^0 \sqrt{\frac{\Delta \Sigma}{\tau}} (\gamma^A \Gamma_A) = i \frac{\Delta}{4r\sqrt{\tau}} \Big( \frac{\partial_r \Delta}{\Delta} + \frac{\partial_r \Sigma}{\Sigma} \Big) \gamma^0 \gamma^1 + i \frac{\sqrt{\Delta}}{4\sqrt{\tau}} \Big( \frac{\partial_\theta \Sigma}{\Sigma} + 2 \cot\theta - 2 \tan\theta \Big) \gamma^0 \gamma^2 + V_0. 
\end{equation}
Hence the Dirac equation becomes
\begin{equation} \label{DE-1}
\begin{split}
  i \partial_t \phi & = \Big[ i\gamma^0 \gamma^1 \frac{\Delta}{r \sqrt{\tau}} \Big( \partial_r + \frac{\partial_r \Delta}{4\Delta} + \frac{\partial_r \Sigma}{4\Sigma} \Big) + i \gamma^0 \gamma^2 \sqrt{\frac{\Delta}{\tau}} \Big( \partial_\theta + 2 \cot\theta - 2 \tan\theta + \frac{\partial_\theta \Sigma}{4\Sigma} \Big) \\
                    & \hspace{1cm} - i\Big( \Omega_a - \gamma^0 \gamma^3 \frac{\sqrt{\Delta \Sigma \beta}}{\tau \sin\theta} \Big) \partial_\varphi - i\Big( \Omega_b + \gamma^0 \gamma^3 \frac{ab\mu\sqrt{\Delta \Sigma}\sin\theta}{\tau \sqrt{\beta}} - \gamma^0 \gamma^5 \frac{\sqrt{\Delta}\Sigma}{\sqrt{\beta \tau}\cos\theta} \Big) \partial_\psi \\
                    & \hspace{2cm} + V_0 + i m \gamma^0 \sqrt{\frac{\Delta \Sigma}{\tau}} \Big] \phi
\end{split}                    
\end{equation}

We now further simplify the equation (\ref{DE-1}) in several steps in order to obtain a synthetic expression adapted to our analysis. We can first get rid of some long range terms appearing in (\ref{DE-1}) by multiplying the spinor by the measure density of the hypersurface $\Sigma = \{t=\textrm{const}\}$ associated to a "good" radial variable. We introduce this new radial variable $x$ such that the principal null geodesics (\ref{PNG}) have radial speed $\pm 1$ with respect to this coordinate. Hence, the "Regge-Wheeler" like coordinate $x$ is defined implicitly by 
\begin{equation} \label{RW-1}
  \frac{dx}{dr} = \frac{(r^2+a^2)(r^2+b^2)}{\Delta} = 1 + \mu \frac{r^2}{\Delta}.
\end{equation}
Using (\ref{Delta}), we can integrate (\ref{RW-1}) to obtain
\begin{equation} \label{RW}
  x = r + \frac{1}{2\kappa_+} \ln \Big( \frac{r-r_+}{r+r_+} \Big) +  \frac{1}{2\kappa_-} \ln \Big( \frac{r-r_-}{r+r_-} \Big) + C,
\end{equation}
where 
$$
\kappa_+ = \frac{r_+(r_+^2-r_-^2)}{(r_+^2+a^2)(r_+^2+b^2)} > 0, \quad \kappa_- = \frac{r_-(r_-^2-r_+^2)}{(r_-^2+a^2)(r_-^2+b^2)} < 0,
$$ 
are called the surface gravities of the event and Cauchy horizons respectively, and $C$ is any constant of integration. Note from (\ref{RW}) that the event horizon $\{r=r+\}$ is now described by $\{x = -\infty\}$. The range of the variable $x$ is then clearly $\R$.  

The measure dvol induced by the metric on the hypersurface $\Sigma = \{t=\textrm{const}\}$ is then 
$$
  \textrm{dvol} \ = \sqrt{\frac{r^2 \Delta \Sigma \tau}{(r^2+a^2)^2 (r^2+b^2)^2}} \textrm{d}x \textrm{d}\omega,
$$  
where d$\omega = \sin\theta \cos\theta$ d$\theta$ d$\varphi$ d$\psi$ is the standard measure on $S^3$. We define the spinor density
$$
  u = \Big( \frac{r^2 \Delta \Sigma \tau}{(r^2+a^2)^2 (r^2+b^2)^2} \Big)^{\frac{1}{4}} \phi,
$$
which satisfies the same equation as the one satisfied by $\phi$ except that $\partial_r$ and $\partial_\theta$ are respectively replaced by 
$$
  \partial_r - \frac{1}{4} \Big( \frac{\partial_r \Delta}{\Delta} + \frac{\partial_r \Sigma}{\Sigma} + \frac{\partial_r h^{-4}}{h^{-4}} \Big), \quad  \quad \partial_\theta - \frac{1}{4} \Big( \frac{\partial_\theta \Sigma}{\Sigma} + \frac{\partial_\theta \tau}{\tau} \Big), 
$$  
where 
\begin{equation} \label{h}
  h^4 = \frac{(r^2+a^2)^2(r^2+b^2)^2}{r^2 \tau}.  
\end{equation}
Hence the Dirac equation becomes
\begin{equation} \label{DE-2}
\begin{split}
  i \partial_t u & = \Big[ i\gamma^0 \gamma^1 \frac{\Delta}{r \sqrt{\tau}} \Big( \partial_r + \frac{\partial_r h}{h} \Big) + i \gamma^0 \gamma^2 \sqrt{\frac{\delta}{\tau}} \Big( \partial_\theta + 2 \cot\theta - 2 \tan\theta - \frac{\partial_\theta \tau}{4\tau} \Big) \\
                    & \hspace{1cm} - i\Big( \Omega_a - \gamma^0 \gamma^3 \frac{\sqrt{\Delta \Sigma \beta}}{\tau \sin\theta} \Big) \partial_\varphi - i\Big( \Omega_b + \gamma^0 \gamma^3 \frac{ab\mu\sqrt{\Delta \Sigma}\sin\theta}{\tau \sqrt{\beta}} - \gamma^0 \gamma^5 \frac{\sqrt{\Delta}\Sigma}{\sqrt{\beta \tau}\cos\theta} \Big) \partial_\psi \\
                    & \hspace{2cm} + V_0 + i m \gamma^0 \sqrt{\frac{\Delta \Sigma}{\tau}} \Big] u.
\end{split}                    
\end{equation}
Noting that
$$
  \frac{\Delta}{r \sqrt{\tau}} \Big( \partial_r + \frac{\partial_r h}{h} \Big) = h \partial_x h, 
$$  
$$  
  \sqrt{\frac{\Delta}{\tau}} \Big( \partial_\theta + \frac{\cot \theta}{2} - \frac{\tan\theta}{2} - \frac{\partial_\theta \tau}{4\tau} \Big) = h \frac{r\sqrt{\Delta}}{(r^2+a^2)(r^2+b^2)} \Big( \partial_\theta + \frac{\cot \theta}{2} - \frac{\tan\theta}{2} \Big) h,
$$
we obtain 
\begin{equation} \label{DE-3}
\begin{split}
  i \partial_t u & = \Big[ h \Big( i\gamma^0 \gamma^1 \partial_x  + i \gamma^0 \gamma^2 \frac{r\sqrt{\Delta}}{(r^2+a^2)(r^2+b^2)} \Big( \partial_\theta + \frac{\cot \theta}{2} - \frac{\tan\theta}{2} \Big) \Big) h \\
                    & \hspace{1cm} - i\Big( \Omega_a - \gamma^0 \gamma^3 \frac{\sqrt{\Delta \Sigma \beta}}{\tau \sin\theta} \Big) \partial_\varphi - i\Big( \Omega_b + \gamma^0 \gamma^3 \frac{ab\mu\sqrt{\Delta \Sigma}\sin\theta}{\tau \sqrt{\beta}} - \gamma^0 \gamma^5 \frac{\sqrt{\Delta}\Sigma}{\sqrt{\beta \tau}\cos\theta} \Big) \partial_\psi \\
                    & \hspace{2cm} + V_0 + i m \gamma^0 \sqrt{\frac{\Delta \Sigma}{\tau}} \Big] u.
\end{split}                    
\end{equation}
Our goal now is to put the equation in the form $i\partial_t u = h \DO h + M(x, \theta, \partial_\varphi, \partial_\psi)$ with $\DO$ a spherically symmetric like Dirac operator and $M$ a short-range (in $r$) first order differential operator in $\partial_\varphi$ and $\partial_\psi$. We thus introduce the Dirac operator on $S^3$ given in our coordinates by
\begin{equation} \label{DS3}
  \DS = i \gamma^0 \gamma^2 \Big( \partial_\theta + \frac{\cot \theta}{2} - \frac{\tan\theta}{2} \Big) + i \gamma^0\gamma^3 \frac{1}{\sin\theta} \partial_\varphi + i \gamma^0\gamma^5 \frac{1}{\cos\theta} \partial_\psi,
\end{equation}
and using the identities
$$
  i m \gamma^0 \sqrt{\frac{\Delta \Sigma}{\tau}} = h \Big( i m \frac{r^2 \sqrt{\Delta}}{(r^2+a^2)(r^2+b^2)} \gamma^0 \Big) h + \ i m \sqrt{\frac{\Delta}{\tau}} (\sqrt{\Sigma} - r) \gamma^0,
$$
$$
  \Omega_a = h \frac{a}{r^2+a^2} h + \frac{ar(\mu r - \sqrt{\tau})}{(r^2+a^2)^2(r^2+b^2)^2}, \quad \quad \Omega_b = h \frac{b}{r^2+b^2} h + \frac{br(\mu r - \sqrt{\tau})}{(r^2+a^2)^2(r^2+b^2)^2},
$$
we get
\begin{equation} \label{DE-4}
\begin{split}
  i \partial_t u & = h \Big[ i\gamma^0 \gamma^1 \partial_x  + \frac{r\sqrt{\Delta}}{(r^2+a^2)(r^2+b^2)} \DS  - \frac{ia}{r^2+a^2} \partial_\varphi - \frac{ib}{r^2+b^2} \partial_\psi + im \frac{r^2 \sqrt{\Delta}}{(r^2+a^2)(r^2+b^2)} \gamma^0 \Big] h u \\
                    & \hspace{1cm} - i\Big( \frac{ar(\mu r - \sqrt{\tau})}{(r^2+a^2)^2(r^2+b^2)^2} - \gamma^0 \gamma^3 \sqrt{\frac{\Delta}{\tau}} \frac{1}{\sin\theta} \Big( \sqrt{\frac{\Sigma \beta}{\tau}} - 1 \Big) \Big) \partial_\varphi u \\
                    & \hspace{1.5cm} - i\Big( \frac{br(\mu r - \sqrt{\tau})}{(r^2+a^2)^2(r^2+b^2)^2} + \gamma^0 \gamma^3 \frac{ab\mu\sqrt{\Delta \Sigma}\sin\theta}{\tau \sqrt{\beta}} - \gamma^0 \gamma^5 \sqrt{ \frac{\Delta}{\tau}}  \frac{1}{\cos\theta} \Big( \frac{\Sigma}{\sqrt{\beta}} - 1 \Big) \Big) \partial_\psi u \\
                    & \hspace{3cm} + V_0 u + i m \gamma^0 \sqrt{\frac{\Delta}{\tau}} (\sqrt{\Sigma} -r) u.
\end{split}                    
\end{equation}

We shall use the following notations. As usual, we denote by $D_x, D_\varphi, D_\psi$ the differential operators $-i \partial_x, \ -i \partial_\varphi, \ -i \partial_\psi$ respectively. We introduce the gamma Dirac matrices $\Go  = i \gamma^0, \quad \Gamma^j = -\gamma^0 \gamma^j, \ j=1,2,3,5$ that satisfy the anticommutation relations
\begin{equation} \label{AntiCommutation}
  \Gamma^a \Gamma^b + \Gamma^b \Gamma^a = 2 \delta_{a,b}.
\end{equation}
Finally we denote the different functions appearing in (\ref{DE-4}) by
\begin{equation} \label{Notations}
\begin{split}
  & a(x)  = \frac{r\sqrt{\Delta}}{(r^2+a^2)(r^2+b^2)}, \quad b(x) = m \frac{r^2 \sqrt{\Delta}}{(r^2+a^2)(r^2+b^2)}, \\
  & c_\varphi(x) = \frac{a}{r^2+a^2}, \quad c_\psi(x) = \frac{b}{r^2+b^2}, \\
  & M_\varphi(x,\theta) = \frac{ar(\mu r - \sqrt{\tau})}{(r^2+a^2)^2(r^2+b^2)^2} + \Gc \sqrt{\frac{\Delta}{\tau}} \frac{1}{\sin\theta} \Big( \sqrt{\frac{\Sigma \beta}{\tau}} - 1 \Big), \\
  & M_\psi(x,\theta) = \frac{br(\mu r - \sqrt{\tau})}{(r^2+a^2)^2(r^2+b^2)^2} - \Gc \frac{ab\mu\sqrt{\Delta \Sigma}\sin\theta}{\tau \sqrt{\beta}} + \Gd \sqrt{ \frac{\Delta}{\tau}}  \frac{1}{\cos\theta} \Big( \frac{\Sigma}{\sqrt{\beta}} - 1 \Big), \\
  & M_0(x,\theta) = V_0(x,\theta) + m \sqrt{\frac{\Delta}{\tau}} (\sqrt{\Sigma} -r) \Go, \\
  & M(x,\theta, D_\varphi, D_\psi) = M_\varphi D_\varphi + M_\psi D_\psi + M_0.     
\end{split} 
\end{equation}
The Dirac equation under Hamiltonian form finally reads 
\begin{equation} \label{HamiltonianD}
\begin{split}
  & i\partial_t u = \DR u, \\
  & \DR = h(x,\theta) \DO h(x,\theta) + M(x,\theta,D_\varphi,D_\psi),
\end{split}   
\end{equation}
where the reference Dirac operator $\DO$ is given by
\begin{equation} \label{HamiltonianDO}
  \DO = \Ga D_x + a(x) \DS + b(x) \Go + c_\varphi(x) D_\varphi + c_\psi(x) D_\psi,
\end{equation}
with $h(x,\theta), \ M(x,\theta, D_\varphi, D_\psi), \ a(x), \ b(x), \ c_\varphi(x), \ c_\psi(x)$ as in (\ref{h}) and (\ref{Notations}).


\section{Spectral properties of the Dirac Hamiltonian} \label{SpectralProperties}

\subsection{Self-adjointness}

In this section, we start studying the basic properties of the Hamiltonians $\DO$ and $\DR$ such as their self-adjointness on a certain domain that we shall describe. 

First of all, let us define precisely the Hilbert space framework for our analysis. Thanks to the weight we have introduced on the spinor, the Hamiltonians $\DO$ and $\DR$ act naturally on the Hilbert space
\begin{equation}
 \H = L^2(\R \times S^3, dx d\omega; \C^4), 
\end{equation}
where  $\omega$ denotes the angular variables $(\theta, \varphi, \psi)$.

To proceed further, we need to know the asymptotic behaviors of the different functions appearing in the definition of $\DO$ and $\DR$. To do so easily, it is useful to have an abstract analytic framework in which the behaviors of these different functions can be read off immediately. Since we work now using the "Regge-Wheeler" variable $x$ and that all the functions in (\ref{Notations}) are expressed in the radial variable $r$, we need the asymptotic behavior of $r$ in function of $x$. From (\ref{RW}), we see that
\begin{equation}
  \begin{split}
    & r \sim x, \quad x \to +\infty, \\
    & r-r_+ = O(e^{2\kappa_+ x}), \quad x \to -\infty. 
  \end{split}
\end{equation}
Note in passing from (\ref{Delta}) that
\begin{equation} \label{DeltaAsymp}
  \begin{split}
    & \Delta(r) \sim x^4, \quad x \to +\infty, \\
    & \Delta(r)= O(e^{2\kappa_+ x}), \quad x \to -\infty. 
  \end{split}
\end{equation}
This leads us to introduce the following classes of symbols for the functions defined on the hypersurface $\Sigma_0 = \R_x \times S^3$. Denoting $(1+x^2)^\half$ by $\x$, we define
\begin{equation}
  S^{m,n} = \{ f \in C^\infty(\Sigma_0), \quad \forall \alpha \in \N, \beta \in \N^3, \ \partial_x^\alpha \partial_\omega^\beta f(x,\omega) \in \left\{ \begin{array}{l} O(\x^{m-\alpha}), \ x \to +\infty, \\ O(e^{-n\kappa_+ x}), \ x \to -\infty, \end{array} \right.\} . 
\end{equation}
\begin{equation}
  S^m = \{ f \in C^\infty(\Sigma_0), \quad \forall \alpha \in \N, \beta \in \N^3, \ \partial_x^\alpha \partial_\omega^\beta f(x,\omega) \in O(\x^{m-\alpha}), \ |x| \to +\infty \}.   
\end{equation}
Note the following obvious properties for these spaces
\begin{equation}
\begin{split}
  (i) & \ S^{m,n} \times S^{p,q} \subset S^{m+p,n+q}, \\
  (ii) & \ \partial_x^\alpha: \ S^{m,n} \longrightarrow S^{m-\alpha,n}, \ \forall \alpha \in \N,\\
  (iii) & \ \partial_\omega^\beta: \ S^{m,} \longrightarrow S^{m,n}, \ \forall \beta \in \N. 
\end{split} 
\end{equation}
To make the link between the $x$ and $r$ variables, we also define for the functions defined on the hypersurface $\Sigma_0 = (r_+,+\infty) \times S^3$ the space
\begin{equation}
  \Pi^m = \{ f \in C^\infty(\Sigma_0), \quad \forall \alpha \in \N, \beta \in \N^3, \ \partial_x^\alpha \partial_\omega^\beta f(r,\omega) \in \left\{ \begin{array}{l} O(\r^{m-\alpha}), \ r \to +\infty, \\ O(1), \ r \to r_+. \end{array} \right\} \}.  
\end{equation}
We shall use the following result whose proof is identical to the one given in \cite{Ha1} (Lemma 9.7.1)
\begin{lemma}
  \begin{equation}
  \begin{split}
    & (i) \ f(r) \in \Pi^m \ \Longrightarrow f(r(x)) \in S^{m,0}, \quad \forall \alpha \in \N^*, \partial_x^\alpha f(r(x)) \in S^{m-\alpha, -2}, \\
    & (ii) \ f(x) \in S^{m,n}, \ g(r) \in \Pi^k, \ \Longrightarrow f(x) g(r) \in S^{m+k, n}. 
  \end{split}
  \end{equation}
\end{lemma}
As a first consequence, we see from (\ref{DeltaAsymp}) that
\begin{equation}
  \Delta \in S^{4,-2}, \quad \sqrt{\Delta} \in S^{2,-1}. 
\end{equation}
Let us now give the asymptotics of the functions appearing in the Hamitonians $\DO$ and $\DR$. We start with the functions $a,b,c_\varphi, c_\psi$ (depending only on the radial variable $x$) used in the definition of $\DO$. We have
\begin{lemma} \label{PotentialsDO}
  \begin{equation}
  \begin{split}
    & (i) \ a(x) \in S^{-1,-1}, \quad \exists \epsilon > 0, \forall j \in \N, \ \Big(a(x)-\frac{1}{x}\Big)^{(j)} \in O(\x^{-1-\epsilon-j}), \ x \to +\infty, \\
                              & \hspace{3.2cm} \exists a_-, \ a(x) - a_- e^{\kappa_+ x} \in O(e^{3\kappa_+ x}), \ x \to -\infty, \\
    & (ii) \ b(x) \in S^{0,-1}, \quad b'(x) \in S^{-2,-1}, \quad (b(x) - m) \in O(\x^{-1}), \ x \to +\infty. \\
    & (iii) \ c_\varphi(x) \in S^{-3,0}, \quad c'_\varphi(x) \in S^{-4,-2}, \quad c_\varphi(x) - \omega_a \in O(e^{2\kappa_+}), \ x \to -\infty, \\
    & (iv) \ c_\psi(x) \in S^{-3,0}, \quad c'_\psi(x) \in S^{-4,-2}, \quad c_\psi(x) - \omega_b \in O(e^{2\kappa_+}), \ x \to -\infty.  
  \end{split}
  \end{equation}
\end{lemma}

Hence the potential $a$ is exponentially decreasing at the event horizon and of Coulomb type at infinity. The potential $b$ is also exponentially decreasing at the event horizon and tends to $m$ at infinity. Note that the difference between $b$ and $m$ is of Coulomb type at infinity. Finally, the potentials $c_\varphi$ and $c_\psi$ decay as $\x^{-3}$ at infinity and tend to the constants $\omega_a = \frac{a}{r_+^2 + a^2}$ and $\omega_b = \frac{b}{r_+^2 + b^2}$ at the event horizon but the differences between $c_\varphi$ and $\omega_a$ and $c_\psi$ and $\omega_b$ decay exponentially at the event horizon. 

We now turn our attention to the function $h(x,\theta)$ and the differential operator $M(x,\theta, D_\varphi, D_\psi)$ appearing in the Hamiltonian $\DR$. We have
\begin{lemma} \label{PotentialsD}
  \begin{equation}
  \begin{split}
    & (i) \ h \geq 1, \quad h-1, \ h^2 - 1 \in S^{-2,-1}, \quad \partial_\theta h \in S^{-2,-2}, \\
    & (ii) \ \forall n,m \in \Z, \quad M_{i,j}(x,\theta,n,m) \in S^{-2,-1}, \ i,j=1,...,4.   
  \end{split}
  \end{equation}
\end{lemma}

The function $h(x, \theta)$ is always greater than $1$ and the difference between $h$ and $1$ is exponentially decreasing at the event horizon and decays as $\x^{-2}$ at infinity. Hence, $h-1$ is short-range at both ends of the exterior region. Also, the differential operator $M(x,\theta, D_\varphi, D_\psi)$ when decomposed onto the angular modes $\{e^{in\varphi} e^{im\psi}\}_{n,m \in \Z}$ is a matrix-valued function whose components are short-range, exponentially decreasing at the event horizon and decaying as $\x^{-2}$ at infinity. 

We shall use Lemmas \ref{PotentialsDO} and \ref{PotentialsD} as follows. Since the exterior region of a $5$D Myers-Perry black hole possesses a bi-azimuthal symmetry with respect to the variables $\varphi$ and $\psi$, it will be enough to study the spectral properties of $\DR$ on each angular modes $\{e^{in\varphi} e^{im\psi}\}_{n,m \in \frac{1}{2} + \Z}$. Denoting by $\DR^{nm}$ and $\DO^{nm}$ the restrictions of $\DR$ and $\DO$ on these angular modes, we find
\begin{equation} \label{H}
  \DR^{nm} = h(x,\theta) \DO^{nm} h(x,\theta) + M(x,\theta,n,m),
\end{equation}
with
\begin{equation} \label{HO}
  \DO^{nm} = \Ga D_x + a(x) \DS + b(x) \Go + nc_\varphi(x) + m c_\psi(x).  
\end{equation}
Note that we kept the full Dirac operator $\DS$ on $S^3$ in the expression of $\DR^{nm}$ and $\DO^{nm}$ since our aim is to compare $\DR$ with a spherically symmetric Dirac operator. The interest in the expressions (\ref{H}) and (\ref{HO}) is that the terms involving $\partial_\varphi$ and $\partial_\psi$ can now be treated as potentials. Moreover, using the decomposition
$$
  \DR^{nm} = \DO^{nm} + (h-1) \DO^{nm} + \DO^{nm}(h-1) + (h-1)\DO^{nm} (h-1) + M(x,\theta,n,m),
$$
and using Lemma \ref{PotentialsD}, the Hamiltonian $\DR$ restricted to each of the angular modes $\{e^{in\varphi} e^{im\psi}\}_{n,m \in \frac{1}{2} + \Z}$ can be viewed as a short-range perturbation of the Hamiltonian $\DO$ restricted to the same angular modes. The latter is the restriction of a spherically symmetric Dirac operator for which we shall be able to use a decomposition onto suitably chosen spherical harmonics to deduce its spectral properties. 

The operators $\DR^{nm}$ and $\DO^{nm}$ will be thus the operators at the heart of our analysis. From now on, we shall assume that these operators act on the full Hilbert space $\H$ as follows: we make the parts of $\DR^{nm}$ and $\DO^{nm}$ involving $\DS$ act on $\H$ in the natural way on $\H$ whereas we still consider the parts of $\DR^{nm}$ and $\DO^{nm}$ involving the angular modes $n$ and $m$ as potentials. To avoid confusion, we denote $H$ and $H_0$ the extension of $\DR^{nm}$ and $\DO^{nm}$ respectively for given $(n,m) \in \frac{1}{2} + \Z$. We shall write concisely
\begin{equation} \label{HamiltonianH-HO}
  H = h(x,\theta) H_0 h(x,\theta) + M(x,\theta), \quad H_0 = \Ga D_x + a(x) \DS + b(x) \Go + c(x), 
\end{equation}
where $c(x) = n c_\varphi(x) + m c_\psi(x)$. It is important to keep in mind that $H_0$ and $H$ only coincide with $\DO^{nm}$ and $\DR^{nm}$ on the particular modes $e^{in\varphi} e^{im\psi}$. In what follows, we shall study the spectral properties of $H_0$ and $H$ and shall be able to obtain equivalent assertions for $\DO$ and $\DR$ simply by restriction to given angular modes.

When dealing with the Hamiltonian $H_0$, we can use a decomposition onto well-chosen spin-weighted spherical harmonics that "diagonalizes" the Dirac operator $\DS$ on $S^3$ to simplify the Hamiltonian. Precisely, we shall use constantly the following result
\begin{lemma} \label{SphericalHarmonics}
The Hilbert space $\H$ can be decomposed onto a Hilbert sum 
  $$
    \H = \displaystyle\bigoplus_{l,n,m \in \mathcal{L}} \H_{lnm}, \quad \mathcal{L} = \big\{(l,n,m), \ \ l \in \N^*, \ (n,m) \in \frac{1}{2} + \Z \big\},
  $$
where the $\H_{lnm}$'s are subpaces of $\H$, isometric to $L^2(\R,dx; \C^4)$, which are let invariant through the action of $H_0$. Moreover, the restrictions of $H_0$ to $\H_{lnm}$ are
$$
  H_0^{lnm} = \DO^{lnm} = \Ga D_x + \lambda_{lnm} a(x) \Gb + b(x) \Go + c(x).
$$
Here, the $\lambda_{lnm}\in \{ \frac{3}{2}, \frac{5}{2}, \frac{7}{2}, ...\}$ are the positive eigenvalues of $\DS$.
\end{lemma}
\textbf{Proof}. See Appendix \ref{Separation}. $\diamondsuit$ \\

We can use this Lemma to prove
\begin{prop} \label{SA-HO}
  The Hamiltonian $H_0$ is self-adjoint on $\H$ with domain
  $$
    D(H_0) = \{ u \in \H, \ \|H_0 u\|^2 < \infty \} = \Big\{ u \in \H, \ u = \sum_{l,n,m \in \mathcal{L}} u_{lnm}, \ \sum_{l,n,m \in \mathcal{L}} \Big( \|u_{lnm}\|^2 + \|H_0^{lnm} u_{lnm}\|^2 \Big) < \infty \Big\}. 
  $$
  As a consequence, $\DO$ is self-adjoint on $\H$ with domain
  $$
    D(\DO) = \{ u \in \H, \ \|\DO u\|^2 < \infty \} = \Big\{ u \in \H, \ u = \sum_{l,n,m \in \mathcal{L}} u_{lnm}, \ \sum_{l,n,m \in \mathcal{L}} \Big( \|u_{lnm}\|^2 + \|\DO^{lnm} u_{lnm}\|^2 \Big) < \infty \Big\}. 
  $$
\end{prop}
\textbf{Proof}. The Hamiltonian $H_0$ being spherically symmetric, it is enough to study $H_0^{lnm}$ on each $\H_{lnm}$. But
$$
  H_0^{lnm} = \Ga D_x + \lambda_{lnm} a(x) \Gb + b(x) \Go + c(x),
$$ 
where the functions $a(x), b(x), c(x)$ are bounded by Lemma \ref{PotentialsDO}. By the Kato-Rellich Theorem, for each $(l,n,m) \in \mathcal{L}$, $H_0^{lnm}$ is self-adjoint on $\H_{lnm}$ with domain 
$$
D(H_0^{lnm}) = \{u_{lnm} \in \H_{lnm}, \ \|H_0^{lnm} u_{lnm}\|^2 < \infty \} = H^1(\R, \C^4).
$$
This gives the the result for $H_0$.  Since $H_0$ and $\DO$ coincide on a given $\H_{nm}$ the restriction of $\H$ to the angular modes $e^{in\varphi} e^{im\psi}$, we deduce that $\DO^{nm}$ is self-adjoint on $\H_{nm}$ with its natural domain. Since this can be done on each $\H_{nm}$, the second assertion is proved. $\diamondsuit$ \\

To study $H$ and $\DR$, we need
\begin{lemma} \label{hDO} 
  1) $h(x, \theta) D(H_0) = D(H_0)$ and $h^{-1}(x,\theta) D(H_0) = D(H_0)$. \\
  2) $h(x, \theta) D(\DO)$ and $h^{-1}(x,\theta) D(\DO) = D(\DO)$. \\  
\end{lemma}
\textbf{Proof}. Let $u \in D(H_0) \subset \H$. Since $h, \partial_x h, \partial_\theta h$ are uniformly bounded according to Lemma \ref{PotentialsD}, we have
\begin{eqnarray*}
  \|H_0 (hu)\| & = & \Big\| \Big( (\Ga D_x + a(x) \Gb D_\theta) h \Big) u + h (H_0 u) \Big\|, \\
               &  \leq & C (\|u\| + \|H_0 u\| ) \ < \ \infty.  
\end{eqnarray*}
Hence $hu \in D(H_0)$. Note that the same argument gives $h^{-1} u \in D(H_0)$.  

Conversely, let $v \in D(H_0)$ et set $u = h^{-1} v$. Then $u \in D(H_0)$ from the argument above and $h u = v$. The first assertion is thus proved. The proof for $\DO$ is clearly the same. We omit it. $\diamondsuit$ \\ 

\begin{coro}
  1) $h H_0 h$ is self-adjoint on $\H$ with domain $D(h H_0 h) = D(H_0)$. \\
  2) $h \DO h$ is selfdjoint on $\H$ with domain $D(h \DO h) = D(\DO)$.
\end{coro}
\textbf{Proof}. The operator $(h H_0 h, D(H_0))$ is clearly symmetric according to Proposition \ref{SA-HO} and Lemma \ref{hDO}. Let us prove that $D((h H_0 h)^*) = D(H_0)$. Suppose that $D((h H_0 h)^*)$ contains strictly $D(H_0)$ et let $v \in D((h H_0 h)^*) \setminus D(H_0)$. Then, there exist $\eta \in \H$ such that
$$
  \kl h H_0 h u, v \er = \kl u, \eta \er, \ \forall u \in D(H_0).
$$
Since the operator of multiplication by $h$ is a bounded self-adjoint operator on $\H$ with inverse $h^{-1}$, we get
$$
  \kl H_0 hu , hv \er = \kl hu, h^{-1} \er, \ \forall u \in D(H_0).
$$
But $\varphi = h u \in D(H_0)$ by Lemma \ref{hDO} and $\psi = h v \notin D(H_0)$ by definition of $v$. Then, there exists $\eta \in \H$ such that
$$
  \kl H_0 \varphi, \psi \er = \kl \varphi, \eta \er, \ \forall \varphi \in D(H_0).
$$
Thus $\psi \in D(H_0)^* = D(H_0)$ which contradicts our assumption. 

The proof for $h \DO h$ is identical. $\diamondsuit$ \\

We now can conclude to the self-adjointness of $H$ and $\DR$ with their natural domains.
\begin{prop} \label{SA-DR}
  1) $H$ is self-adjoint on $\H$ with domain $D(H) = \{ u \in \H, \ \|H u\|^2 < \infty \} = D(H_0)$. \\
  2) $\DR$ is self-adjoint on $\H$ with domain 
  $$
    D(\DR) = \{ u \in \H, \ \|\DR u\|^2 < \infty \} = \Big\{ u \in \H, \ u = \sum_{l,n,m \in \mathcal{L}} u_{lnm}, \ \sum_{l,n,m \in \mathcal{L}} \Big( \|u_{lnm}\|^2 + \|\DR^{lnm} u_{lnm}\|^2 \Big) < \infty \Big\}. 
  $$
\end{prop}
\textbf{Proof}. The proof is obvious for $H$ by the Kato-Rellich Theorem and the fact that the potential $M(x,\theta,n,m)$ is bounded for fixed $(n,m) \in \frac{1}{2} + \Z$. Hence, we conclude that $\DR^{nm}$ is self-adjoint on $\H_{nm}$ the restriction of $\H$ to the angular modes $\e^{in\varphi} e^{im\psi}$ with its natural domain. Since this can be done for each $\H_{nm}$, $n,m \in \frac{1}{2} + \Z$, the second assertion follows. $\diamondsuit$ \\

Finally, thanks to Proposition \ref{SA-DR}, the solutions of the Dirac equation (\ref{HamiltonianD}) can be written using the evolution $e^{-it\DR}$ as $u(t) = e^{-it\DR} u_0$ where $u_0 \in \H$ is any initial data on the hypersurface $\Sigma_0$.


\subsection{Local energy decay I}

We begin this section by collecting some standard results useful for the later Mourre analysis and for our first version of local energy decay below. We first give a compactness criterion for operators of the form $f(x) g(H_0)$ as well as $f(x) g(H)$ where $f,g$ are functions decaying at infinities. To do so, we need a better description of the domain $D(H_0) = D(H)$. By the same argument as in \cite{HaN}, we know that
\begin{equation}
 D(H) = D(H_0) \subset H^1_{\textrm{loc}}(\Sigma_0; \C^4).
\end{equation}
Then we have the standard compactness criterion
\begin{lemma} \label{Compactness}
 If $f = f_{ij}, g \in C_\infty(\R)$, then the operator $f(x) g(H_0)$ and $f(x) g(H)$ are compact on $\H$.
\end{lemma}
As a consequence of Lemma \ref{Compactness} and of the Helffer-Sj\"ostrand formula \cite{Dav}, we also have
\begin{coro} \label{CompactnessH-HO}
 If  $\chi \in \cor$, then the operator $\chi(H) - \chi(H_0)$ is compact. 
\end{coro}

We shall also need in the next section the following resolvent estimates whose proof is identical to the one in \cite{HaN}. There exists a constant $C$ such that for any $u \in D(H_0)$, 
\begin{equation} \label{ResolventEstimates}
 \begin{split}
  & \| a(x) \DS u \| \leq C ( \|H_0 u\| + \|u\|), \\
  & \|\Ga D_x u \| \leq C ( \|H_0 u\| + \|u\|).
 \end{split}
\end{equation}
The same estimates remain true if we replace $H_0$ by $H$.  

At last, we state an important property of domain invariance for Dirac operators only (see \cite{Th}). Precisely, the domains $D(\x^n), \ n \in \N$ are stable under the action of the resolvents of $H_0$ and $H$.
\begin{lemma} \label{DomainInvariance}
  Let $n \in \N, \ z \in \C \setminus \sigma(H)$ and $\chi \in \cor$. Then
  \begin{equation}
  \begin{split} \label{DI}
    & (H-z)^{-1} D(\x^n) \subset D(\x^n), \\
    & \chi(H) D(\x^n) \subset D(\x^n). 
  \end{split}
  \end{equation}  
  Moreover, the estimates (\ref{DI}) remain true if we replace $H$ by $H_0$.
\end{lemma}

We now start our study of the spectral properties of the Hamiltonian $\DR$. From Appendix \ref{Separation}, we know that the spectrum of $\DR$ is purely continuous, \textit{i.e.} $\sigma_{\textrm{pp}}(\DR) = \emptyset$. This fact together with the compactness criterion above entails 
\begin{prop} \label{NoPointSpectrum}
  Let $\chi \in \cor$. Then, for all $u \in \H$,
  $$
    \frac{1}{2T} \int_{-T}^{T} \| \chi(x) e^{-it\DR} u \|^2 dt \ \to \ 0.
  $$
\end{prop}
\textbf{Proof}. By a density argument, it is enough to prove this result for $\DR^{nm} = H^{nm}$, the restriction of $\DR$ onto each of the angular modes $\{e^{in\varphi} e^{im\psi} \}$, $n,m \in \frac{1}{2} + \Z$. But, the spectrum $H^{nm}$ is clearly also purely continuous and from Lemma \ref{Compactness}, the operator $\chi(x) (H^{nm} +i)^{-1}$ is compact. Hence, the result follows directly from the RAGE Theorem (see for instance \cite{RS}). $\diamondsuit$ \\


\Section{Mourre theory} \label{Mourre}

In this section, we prove the absence of singular continuous spectrum for the Hamiltonian $H$ defined in (\ref{HamiltonianH-HO}) by means of an application of the Mourre theory. Since this Hamiltonian coincide on each angular modes $\{e^{in\varphi} e^{im\psi}\}$, $n,m \in \frac{1}{2} + \Z$, with the Hamiltonian $\DR^{nm}$, we shall conclude that the spectrum of $\DR$ also contains no singular continuous spectrum and thus, is purely absolutely continuous by Proposition \ref{NoPointSpectrum}. Using this fact, we prove the weak local decay energy stated in Theorem \ref{MainThm} by a simple application of the Riemann-Lebesgue Lemma.   

We begin this section recalling the basics of the Mourre theory. To determine then a locally conjugate operator for $H$, we use the construction of \cite{Da4} where the same problem was studied in the case of the $4$-dimensional Kerr-Newman metric. It turns out that the expression of the Hamiltonian $H$ given in (\ref{HamiltonianH-HO}) enters exactly in the analytic framework studied in \cite{Da4}, except that the Dirac operator of the $2$D-sphere $S^2$ is now replaced by the Dirac operator $\DS$ on the $3$D-sphere. For the convenience of the reader, we reproduce the proof with the slight necessary modifications due to the dimension.

\subsection{Abstract Mourre theory}

In this section, $H$ denotes any self-adjoint operator on a Hilbert space $\H$. Mourre theory is a powerful tool to study the spectral nature of $H$. Its principle consists in finding a self-adjoint operator $A$ on $\H$ so that the pair $(H,A)$ satisfies a set of assumptions which we now recall (see \cite{M}). 
\begin{description}
  \item [(M1)] $e^{-itA} D(H) \subset D(H)$.
  \item [(M2)] $i[H,A]$ defined as a quadratic form on $D(H) \cap
  D(A)$ extends to an element of $\B(D(H),\H)$.
  \item [(M3)] $[A,[A,H]]$ well defined as a quadratic form on
  $D(H) \cap D(A)$ by \textbf{(ii)}, extends to an element of
  $\B(D(H),D(H)^{*})$. 
  \item [(M4)] Let $I \subset \R$ an open interval.  There exists a strictly positive constant $\mu$  and a compact operator $K$ such that 
  \begin{equation}\label{MourreEstimate}
    \mathbf{1}_I(H) i[H,A] \mathbf{1}_I(H) \geq \e \mathbf{1}_I(H) + K.    
  \end{equation}
\end{description}
The fundamental assumption here is the Mourre estimate (\ref{MourreEstimate}). Its meaning is that we must find an observable $A$ which essentially increases along the evolution $e^{-itH}$. The other conditions are more technical and turn out to be difficult to check directly in the case where $A$ and $H$ are unbounded self-adjoint operators having no explicitly known domains. We give below some useful criteria to verify them. If the pair $(H,A)$ satisfy these assumptions then we say
that $A$ is a \emph{conjugate operator} for $H$ on $I$. The existence of a conjugate operator provides important informations on the spectrum of $H$. Precisely, we have (see for example \cite{ABG1})     
\begin{thm} \label{KN-MO}
  Let $H,A$ be self-adjoint operators on $\H$. Assume that $A$ is a
  conjugate operator for $H$ on the interval $I$. Then $H$ has no
  singular continuous spectrum in $I$. Furthermore, the number of
  eigenvalues of $H$ in $I$ is finite (counting
  multiplicity).  
\end{thm}

Let us now give some details concerning the conditions (M1), (M2) and (M3) of Mourre theory. One of the difficulties in Mourre theory consists in working with commutators $i[H,A]$ (see (M2)) between unbounded self-adjoint operators. One has to be careful when dealing with such quantities since $D(H)$ and $D(A)$ may be unknown or have an intersection which is not even dense in $\H$. Similarly, the assumption (M1) is not easy to prove since the action of $e^{isA}$ may also be unknown. Therefore, it is useful to have a different set of criteria. Let us first define the class of operators
$C^k(A)$ introduced in \cite{ABG1}.  
\begin{definition}
  Given a self-adjoint operator $A$, we say that a self-adjoint
  operator $H$ belongs to $C^{k}(A), \ k \in \N,$ if and only if 
  \begin{displaymath}
  \mathbf{(ABG)} \hspace{2cm} \exists z \in \C \setminus \sigma(H),
  s \longrightarrow e^{isA}(H-z)^{-1}e^{-isA} \in
  C^{k}(\R_{s};\mathcal{B}(\H)), \hspace{3cm}
  \end{displaymath}
for the strong topology of $\B(\H)$.
\end{definition}
It is shown in \cite{ABG1} that one can replace the assumptions (M1) and (M2) by the unique assumption $H \in C^1(A)$ without changing the conclusions of Theorem \ref{KN-MO}. Roughly speaking, this condition allows for the following equality
\begin{displaymath}
  [A,(z-H)^{-1}] = (z-H)^{-1} [A,H] (z-H)^{-1}, 
\end{displaymath}
to make sense on $\H$. From \cite{ABG1}, $H \in C^1(A)$ is equivalent
to    
\begin{eqnarray*} 
  \mathbf{(i)'} &  \exists z \in \C \setminus \sigma(H),
  (H-z)^{-1}D(A) \subset D(A), (H-\overline{z})^{-1}D(A) \subset D(A),
  \\
  \mathbf{(ii)'} & |(Hu,Au)-(Au,Hu)| \leq C(\|Hu\|^{2}+\|u\|^{2}),
  \quad \forall u \in D(H) \cap D(A),
\end{eqnarray*}
which is similar to the conditions (M1) and (M2). Nevertheless, they remain complicated to check when the domains of $H$ and $A$ are not explicitly known. One way to remedy this problem consists in finding first a common core for $H$ and $A$. This procedure is described in \cite{GL}. We only recall the two results we shall need.   
\begin{lemma}[Nelson]\label{KN-Nelson}
  Let $N \geq 1$ a self-adjoint operator on $\H$. Let $A$ a
  symmetric operator on $\H$ such that $D(N) \subset
  D(A)$. Assume that
  \begin{equation} 
    \begin{array}{cc}
      \|Au\| \leq C \|Nu\|, \quad \forall u \in D(N),\\
      |(Au,Nu) - (Nu,Au)| \leq C \|N^{\frac{1}{2}}u\|^{2}, \quad
         \forall u \in D(N).\\ 
    \end{array}
  \end{equation}
  Then $A$ is essentially self-adjoint on $D(N)$. Furthermore every
  core of $N$ is also a core for $A$.
\end{lemma}
\begin{lemma}[G\'erard, Laba] \label{KN-GerardLaba}
  Let $H$, $H_{0}$ and $N$ three self-adjoint operators on $\H$
  satisfying $N \geq 1$, $D(H) = D(H_{0})$ and $(H-z)^{-1} D(N) \subset
  D(N)$. Let $A$ a symmetric operator on $D(N)$. Assume that $H_0$ and
  $A$ satisfy the assumptions of Lemma \ref{KN-Nelson} and 
  \begin{equation}
    |(Au,Hu) - (Hu,Au)| \leq C (\|Hu\|^{2} + \|u\|^{2}), \quad \forall
    u \in D(N).
  \end{equation}
  Then we have
  \begin{description}
    \item [$\bullet$] $D(N)$ is dense in $D(A) \cap D(H)$ with the
    norm $\|Hu\|+\|Au\|+\|u\|$,
    \item [$\bullet$] The quadratic form $i[H,A]$ defined on  $D(A)
    \cap D(H)$  is the unique extension of $i[H,A]$ on $D(N)$,
    \item [$\bullet$] $H \in C^{1}(A)$.
  \end{description}
\end{lemma}
The operator $N$ above is called a \emph{comparison operator}. When the assumptions of Lemma \ref{KN-GerardLaba} are satisfied, it is enough to compute the commutator $i[H,A]$ as a quadratic form on the common core $D(N)$ or on any core of $N$. We finish this brief presentation of Mourre theory by an interesting observation. The condition $H \in C^1(A)$ together with the condition (M2) imply the condition (M1) thanks to a result due to G\'erard and Georgescu \cite{GG}
\begin{lemma} \label{KN-GeorgescuGerard}
  Let $H$ and $A$ two self-adjoint operators such that $H \in
  C^{1}(A)$ and $i[H,A] \in \B(D(H),\H)$ then $e^{isA} D(H) \subset
  D(H)$ for all $s \in \R$.
\end{lemma}

The choice of a locally conjugate operator for Hamiltonians $H$ like (\ref{HamiltonianH-HO}) is not straightforward. Recall indeed that the evolution described by $H$ can be understood as an evolution on a Riemannian manifold (given here by $\Sigma_0 = \R_x \times S^3$) having two different ends. At infinity, the metric on $\Sigma_0$ tends to the flat metric. Therefore, we can use the usual generator of dilations there. At the event horizon, this metric is exponentially large and the choice of a conjugate operator turns out to be much more complicated. Analogous situations have been studied before, first by Froese and Hislop \cite{FH} in the case of a second-order-elliptic Hamiltonian, then by De Bi\`evre, Hislop and Sigal
\cite{DHS} for the wave equation on classes of non-compact Lorentzian manifolds with several asymptotic ends which are perturbations of particular simple geometries and more recently by Bouclet \cite{Bo} for the Laplacian on asymptotically hyperbolic manifolds. Closer to our model, H\"afner and Nicolas \cite{HaN} treated the case of massless Dirac fields in a Kerr background, work that was generalized to massive Dirac fields in a Kerr-Newman background in \cite{Da4}. The latter two models are in fact almost identical to ours and we shall use here the construction of a locally conjugate operator given there.

\subsection{Locally conjugate operator for $\DR$}

\subsubsection{Preliminaries}

In order to separate the problems at the event horizon and infinity, we define two cut-off functions $j_\pm \in C^\infty(\R)$ satisfying  
\begin{eqnarray*}
  j_-(x) = 1, \ \textrm{for} \ x \leq \half, \quad j_-(x) = 0, \
  \textrm{for} \ x \geq 1, \\ 
  j_+(x) = 1, \ \textrm{for} \ x \geq 2, \quad j_+(x) = 0, \
  \textrm{for} \ x \leq \frac{3}{2}. 
\end{eqnarray*}
In this definition, the support of $j_-$ must contain $(-\infty,0]$ whereas the support of $j_+$ only has to contain a neighbourhood of $+\infty$. Now, for $S \geq 1$, we set 
\begin{displaymath}
  R_-(x,\DS) = (x + \kappa_+^{-1} \ln |\DS|) \, j_-^2 \Big(
  \frac{x + \kappa_+^{-1} \ln |\DS|}{S} \Big), \quad \quad R_+(x)
  = x j_+^2(\frac{x}{S}).  
\end{displaymath}
We define the local conjugate operators $A_-,A_+$ by
\begin{displaymath}
  A_- = R_-(x,\DS) \Ga, \quad \quad A_+ = \half (D_x \, R_+(x) + \, R_+(x) D_x).  
\end{displaymath}
At infinity, the conjugate operator $A_+$ is like a generator of dilations whereas at the horizon, $A_-$ is the locally conjugate operator introduced in \cite{Da4}. The true conjugate operator $A$ will be the sum or the difference between $A_-$ and $A_+$ depending on the energy interval we consider. Precisely, let $\chi \in \cor$ such that $supp\,\chi$ is included in one of the intervals $(-\infty,-m)$, $(-m,+m)$, $(+m,+\infty)$. Then we define the conjugate operator $A_\chi$ on $supp\,\chi$ by   
\begin{eqnarray*}
  A_\chi & = & A_- + A_+ \ \textrm{when} \ supp\,\chi \subset
  (+m,+\infty), \\
  A_\chi & = & A_- - A_+ \ \textrm{when} \ supp\,\chi \subset
  (-\infty,-m), \\
  A_\chi & = & A_- \ \textrm{when} \ supp\,\chi \subset (-m,+m). 
\end{eqnarray*}
We will prove the assumptions (M1) to (M3) for $A = A_- + A_+$. The other cases are analogous. 

Before starting with the proof, we simplify some notations. In what follows indeed, we shall use extensively the decomposition onto generalized spherical harmonics for the Dirac operator $\DS$ introduced in Lemma \ref{SphericalHarmonics}. Recall that on each Hilbert space $\H_{lnm}$, the operator $H_0$ reduces to 
$$
  H_0^{lnm} = \Ga D_x + \lambda_{lnm} a(x) \Gb + b(x) \Go + c(x), 
$$
where $\lambda_{lnm} \in \{ \frac{3}{2}, \frac{5}{2}, \frac{7}{2}, ...\}$ are the positive eigenvalues of $\DS$. Since $H_0^{lnm}$ only depends on $\lambda_{lnm}$ and for ease of notations, we shall gather all the indices $(l,n,m)$ corresponding to the same eigenvalues $\lambda_{lnm} = k$. Recalling that there exists only a finite number of combinations of $(l,n,m)$ such that $\lambda_{lnm} = k$, we thus introduce in a natural way a new decomposition of the Hilbert space $\H$, in the form 
$$
  \H = \bigoplus_{k \in \frac{3}{2} + \N} \H_k, \hspace{1cm} \H_k = \bigoplus_{(l,n,m) / \ \lambda_{lnm} = k} \H_{lnm}, \hspace{1cm} \H_k \simeq L^2(\R;\C^4),
$$
such that 
$$
  H_0^k = H_{0 \,|\H_k} = \Ga D_x + k a(x) \Gb + b(x) \Go + c(x). 
$$

Let us now study the operator $R_-(x,\DS)$ appearing in $A_-$. On each Hilbert space $\H_k$, it reduces to multiplication operator $R_-(x, k)$ where $k \in \frac{3}{2} + \N$. We have 
\begin{lemma} \label{KN-CO-1}
  For all $S \geq 1$,
  \begin{eqnarray*}
    |R_-(x,k)| \leq C \x, \ \textrm{uniformly in k}, \\
    |R_-^{(i)}(x,k)| \leq C, \ \textrm{uniformly in k}, \ i=1,2.   
  \end{eqnarray*}
  As a consequence, the domains of the operators $R_-(x,\DS)$ and $A_-$ contain $D(\x)$. Moreover, if $j_1 \in C^\infty(\R)$ satisfies $j_1(x) = 1$, for $x \leq 1$ and $j_1(x) = 0$, for $x \geq \frac{3}{2}$, then  
  \begin{displaymath}
    R^{(i)}_-(x, \DS) = R^{(i)}_-(x,\DS) j_1^2(\frac{x}{S}), \
    i= 0,1,2. 
  \end{displaymath}
\end{lemma}
\textbf{Proof}. The operator $R_-(x,k)$ is a translation by $\kappa_+^{-1} \ln |k|$ of the operator of multiplication $x j_-^2(\frac{x}{S})$. Since $|k| \geq \frac{3}{2}$, we have 
\begin{equation} \label{KN-CO-1a}
  x \leq x + \kappa_+^{-1} \ln |k|, \quad \forall k.
\end{equation}
Therefore, on $supp\,j_-$ we have
\begin{displaymath}
  x \leq x + \kappa_+^{-1} \ln |k| \leq S, 
\end{displaymath}
and thus
\begin{displaymath}
  |x + \kappa_+^{-1} \ln |k|| \leq \max(S,|x|) \leq C \x, \quad
   \forall k. 
\end{displaymath}
This proves the first assertion. 

Now we have
\begin{displaymath}
  R_-^{(1)}(x,k) = j_-^2 \Big( \frac{x + \kappa_+^{-1} \ln
  |k|}{S} \Big) + \Big( \frac{x + \kappa_+^{-1} \ln |k|}{S}
  \Big) (j_-^2)^{'} \Big( \frac{x + \kappa_+^{-1} \ln |k|}{S}
  \Big). 
\end{displaymath}
Since $supp\, (j_-^2)^{'}$ is compact, the second assertion holds. 

Finally, it is immediate from (\ref{KN-CO-1a}) that 
\begin{displaymath}
  (j_-^2)^{(i)} \Big( \frac{x + \kappa_+^{-1} \ln |k|}{S} \Big)
  j^2_1(\frac{x}{S}) = (j_-^2)^{(i)} \Big( \frac{x + \kappa_+^{-1}
  \ln |k|}{S} \Big), \quad \forall k, \ i=0,1,2.  
\end{displaymath}
This concludes the proof of the lemma. $\diamondsuit$ \\

We shall now prove that the Mourre assumptions (M1) to (M3) hold for $(H,A)$. We first define the comparison operator
\begin{displaymath}
  N = D_x^2 + a^2(x) \DS^2 + \x^2, 
\end{displaymath}
where $a(x)$ is the potential in front of $\DS$ in the definition of $H$. By the same proof as in \cite{HaN}, we can characterize the domain of $N$ by $D(N) = D(H^2) \cap D(\x^2) = D(H_0^2) \cap D(\x^2)$. From this and Proposition \ref{DomainInvariance}, we have for any $z \in \C \setminus \sigma(H)$, 
\begin{equation} \label{DI-N-H}
  (H-z)^{-1} D(N) \subset D(N).
\end{equation}
We also state some useful estimates (immediate from the definition of $N$): for $u \in D(N)$,  
\begin{equation} \label{KN-COp}
  \|D_x u \| \leq C \|N^\half u\|, \quad \|a(x) \DS u\| \leq C
  \|N^\half u\|, \quad \|x u\| \leq C \|N^\half u \|.  
\end{equation}
Our first result shows that the pairs of Hamiltonians $(A_\pm,N)$ and $(H_0,N)$ satisfy the hypotheses of Lemma  \ref{KN-Nelson} above. Precisely, we have
\begin{lemma} \label{KN-CO-2}
  $(A_\pm,N)$ and $(H_0,N)$ satisfy the hypotheses of Lemma \ref{KN-Nelson}. In particular, the operators $A_\pm$ and $H_0$ are essentially self-adjoint on $D(N)$. 
\end{lemma}
\textbf{Proof}. We first consider the pair $(A_\pm,N)$. By Lemma \ref{KN-CO-1}, $D(N) \subset D(\x) \subset D(A_-)$ and for any $u \in D(N)$, we have
\begin{displaymath}
  \|A_- u\| \leq C\|\x u\| \leq C \|Nu\|. 
\end{displaymath}
Hence, it remains to show that $|(u,i[A_\pm,N]u)| \leq C \|N^\half u\|^2$. But, using Lemma \ref{KN-CO-1} and (\ref{KN-COp}), we have
\begin{displaymath}
  |(u,[A_-,N]u)| \leq 2 |(D_x u, R_-^{(1)}(x,\DS) \Ga u)| \leq C
   \|u\| \|D_x u\| \leq C\|N^\half u\|^2. 
\end{displaymath}
The proof for $A_+$ is identical to that given in \cite{HaN}. We omit it. 

Let us now consider the pair $(H_0,N)$. Using (\ref{KN-COp}), we have for any $u \in D(N)$
\begin{displaymath}
  \|H_0 u\| \leq \|D_x u\| + \| a(x) \DS u\| + C \|u\| \leq C
  \|Nu\|. 
\end{displaymath}
Moreover, 
\begin{eqnarray*}
  |(u,[H_0,N]u)| & \leq & 2 \Big( |(u,\Ga a^{'}(x) a(x) \DS^2
   u)| + |(u, \Ga x u)| + |(D_x u, a^{'}(x) \DS u)| \\
                 &      & + |(D_x u, b^{'}(x) u)| + |(D_x u,
   c^{'}(x) u)| \Big).   
\end{eqnarray*}
Now remark that there exists a constant $C$ such that $|a^{'}(x)|
\leq C |a(x)|$. Hence we have
\begin{eqnarray*}
  |(u,[H_0,N]u)| & \leq & C \Big( \|a(x) \DS u\|^2 + \|u\|\, \|x
                 u\| + \|D_x u\| (\| a(x) \DS u\| + \|u\| ) \Big),
                 \\  
                 & \leq & C \|N^\half u\|^2.
\end{eqnarray*}
This proves the assertion. $\diamondsuit$\\

\begin{lemma} \label{KN-CO-4}
  $H \in C^1(A)$. Moreover, the commutator $i[H,A]$ belongs to  $\B(D(H), \H)$. It follows that the assumptions (M1) and (M2) of the Mourre theory are satisfied.
\end{lemma}
\textbf{Proof}: Thanks to Lemma \ref{KN-CO-2}, it suffices to show that $|(u,i[H,A_\pm]u)| \leq C (\|H u\|^2 + \|u\|^2)$ for all $u \in D(N)$ in order to apply Lemma \ref{KN-GerardLaba}. We only prove it for $A_-$ since the proof for $A_+$ is identical to that given in \cite{HaN}. We first calculate $i[H_0,A_-]$. We have 
\begin{displaymath}
  i[H_0, A_-] = R_-^{(1)}(x,\DS) + 2i a(x) R_-(x,\DS) \DS \Ga
  + 2i b(x) R_-(x,\DS) \Go \Ga. 
\end{displaymath}
Let $u \in D(N)$. By Lemma \ref{KN-CO-1}, we estimate the first term by $\|R_-^{(1)}(x,\DS) u\| \leq C \|u\|$. We can estimate the second term by  
\begin{equation}
\begin{split} 
  \|a(x) R_-(x,\DS) \DS \Ga u \| & \leq \| \frac{a(x) - a_-
   e^{\kappa_+ x}}{a(x)} j^2_1(\frac{x}{S}) R_-(x,\DS) \| \|
   a(x) \DS (H_0+i)^{-1}\| \|(H_0+i)u\| \\  
                                 &  \hspace{2cm} + \|a_- R_-(x,\DS) e^{\kappa_+(x + \kappa_+^{-1} \ln|\DS|)} 
   \frac{\DS}{|\DS|} \Ga u \|.
\end{split}   
\end{equation}
Observe that $\| \frac{a(x) - a_- e^{\kappa_+ x}}{a(x)} j^2_1(\frac{x}{S}) R_-(x,\DS) \|$ is bounded thanks to Lemmas \ref{PotentialsDO} and \ref{KN-CO-1}. Furthermore, $\| a(x) \DS (H_0+i)^{-1}\|$ and $\|R_-(x,\DS) e^{\kappa_+(x + \kappa_+^{-1} \ln|\DS|)} \|$ are bounded by the resolvent estimates (\ref{ResolventEstimates}) for the former and by definition of $R_-(x,\DS)$ for the latter. Hence, we get
\begin{displaymath}
  \|a(x) R_-(x,\DS) \DS \Ga u \| \leq C(\|H_0 u\| + \|u\|).
\end{displaymath}
Finally, the last term is estimated as follows 
\begin{displaymath}
  \|b(x) R_-(x,\DS) \Go \Ga u\| \leq C \|b(x)
  j^2_1(\frac{x}{S}) R_-(x,\DS)\| \|u\| \leq C \|u\|,
\end{displaymath}
again by Lemma \ref{PotentialsDO}. In summary, we have 
$$
  i[H_0,A_-] \leq C ( \|H_0 u\| + \|u\| ).
$$ 

We now estimate $i[H,A_-]$. We have $i[H,A_-] = h \,i[H_0,A_-] h + i[M,A_-]$. Since $h$ is bounded as an operator from $D(H) = D(H_0)$ into itself, the first term belongs to $\B(D(H), \H)$ by the previous estimate. Moreover, since $M \in S^{-2}(\R)$ according to Lemma \ref{PotentialsD}, the remaining term $i[M,A_-]$ is clearly bounded by Lemma \ref{KN-CO-1}. This concludes the proof of the Lemma. $\diamondsuit$ \\

\begin{lemma} \label{KN-CO-5}
  The double commutator $[i[H,A],A]$ extends to a bounded operator in $\B(D(H), \H)$. 
\end{lemma}
\textbf{Proof}: We first estimate $[i[H,A_-],A_-]$. We have
\begin{displaymath}
  [i[H,A_-],A_-] = h [i[H_0,A_-],A_-]h + [i[M,A_-],A_-]. 
\end{displaymath}
The second term is clearly bounded since $M \in S^{-2}$ and $A_-$ is bounded from $D(\x)$ to $\H$. Moreover, the first term
\begin{displaymath}
  [i[H_0,A_-],A_-] = 4ia(x) R^2_-(x,\DS) \DS + 4ib(x)
  R^2_-(x,\DS) \Go,  
\end{displaymath}
is bounded from $D(H)$ to $\H$ by the same argument as in the proof of Lemma \ref{KN-CO-4}. 

Now we estimate $[i[H,A_-],A_+]$. We have
\begin{equation} \label{KN-CO-5a}
  [i[H,A_-],A_+] = \Big( i[h,A_+] [H_0,A_-] h + h.c. \Big) +  h[i[H_0,A_-],A_+] h + [i[M,A_-],A_+], 
\end{equation}
where $h.c.$ denotes the hermitian conjugate of the quantity on its left. Since $i[h,A_+] = -R_+(x) (\partial_{x} h) \in S^{-1}$ by Lemma \ref{PotentialsD}, the first term in (\ref{KN-CO-5a}) is bounded from $D(H)$ to $\H$ by Lemma \ref{KN-CO-4}. From the exact expression of $i[H_0,A_-]$, the second term in (\ref{KN-CO-5a}) is written as
\begin{equation} \label{KN-CO-5b}
\begin{split}
  [i[H_0,A_-],A_+] & = iR_+(x) R_-^{(2)}(x,\DS) - 2 \big( a(x) R_-(x,\DS) \big)^{(1)} R_+(x) \DS \Ga \\
                   & \hspace{2cm} -2 \big( b(x) R_-(x,\DS) \big)^{(1)} R_+(x) \Go.
\end{split}                    
\end{equation}
Recall that $R^{(i)}_-(x,\DS) = R^{(i)}_-(x,\DS) j_1^2(\frac{x}{S})$, $\forall i=0,1,2$ by Lemma \ref{KN-CO-1}. We can assume that the function $j_+$ has been chosen such that $supp\, j_1 \cap supp\,j_+ = \emptyset$. Hence, (\ref{KN-CO-5b}) vanishes. Finally, since $[M,A_-] \in S^{-1}$, the last term in (\ref{KN-CO-5a}) is also bounded from $D(H)$ to $\H$.    

Similarly, $[i[H,A_+],A_-]$ is bounded from $D(H)$ to $\H$. The same result holds for $[i[H,A_+],A_+]$ (see
\cite{HaN}). $\diamondsuit$\\  

In conclusion, we have proved that the pair of self-adjoint operators $(H,A)$ satisfies the technical assumptions (M1), (M2) and (M3) of Mourre theory. It remains to check that the Mourre estimate (\ref{MourreEstimate}) holds on suitably chosen intervals.


\subsubsection{The Mourre estimate for $(H,A)$}

The strategy in this section is the following. We first establish Mourre estimates between $H_0$ and $A_-$ (resp. $H_0$ and $A_+$). The main difficulties arise with the part of the proof concerned with $A_-$. We shall use here several technical results from \cite{HaN} which hold true in our model. Since they are instructive, short proofs will be sketched in the course of the calculations. The full results can be found in Section 5.5 of \cite{HaN}. Finally, we will show that the remaining term involving $H - H_0$ is compact on $\H$ and thus is negligible (in the sense of Mourre theory). \\ 

\noindent \textbf{Mourre estimate for $(H_0,A_-)$}: When working at the horizon of the black hole, it is convenient to introduce the operator 
\begin{displaymath}
  H_e = \Ga D_x + a_- e^{\kappa_+ x} \DS + c_-, \hspace{1cm} c_- = n \omega_a + m \omega_b, 
\end{displaymath}
which corresponds to the formal limit of the operator $H_0$ when $x \to -\infty$ (see (\ref{HamiltonianH-HO}) and Lemma \ref{PotentialsDO}). Let us recall some basic properties of $H_e$ (see Sections 3.2 and 5.5 in \cite{HaN}). On each reduced Hilbert space $\H_k$ with $k \in \frac{3}{2} + \N$, we denote this operator $H_e^k$ and we have $D(H_e^k) = \{ u \in \H_k, \, H_e^k u \in \H_k\}$ its natural domain. Then the following properties hold 
\begin{eqnarray}
  \textrm{Resolvent estimates}: & \forall u \in D(H_e^k), \
  \max(\|\Ga D_x u\|, \|k\, e^{\kappa_+ x} \Gb u\|) \leq C (\|H_e^k  u\| + \|u\|), \label{KN-ResEst-3}\\
  \textrm{self-adjointness}: & (H_e^k, D(H_e^k)) \ \textrm{is self-adjoint on} \ \H_k, \\
  \textrm{Characterization of the domain}: & D(H_e^k) \subset [H^1(\R)]^4, \\
  \textrm{Compactness criterion}: & \forall f,g \in C_\infty(\R), \
  f(x) g(H_e^k) \ \textrm{is compact}, \label{KN-Comp0}\\
  \textrm{Spectrum}: & H_e^k \ \textrm{has no eigenvalue} \ i.e. \ \sigma_{pp}(H_e^k) = \emptyset. \label{KN-Eigen}
\end{eqnarray}
We also denote $D(H_e) = \{ u = \sum u_k, \, u_k \in D(H_e^k), \sum (\|H_e^k u\|^2 + \|u\|^2) < \infty \}$ the domain of $H_e$. We have
\begin{eqnarray}
  \textrm{Resolvent estimates}: & \forall u \in D(H_e), \
  \max(\|\Ga D_x u\|, \|e^{\kappa_+ x} \DS u\|) \leq C (\|H_e u\| +
  \|u\|), \label{KN-ResEst-4}\\ 
  \textrm{self-adjointness}: & (H_e, D(H_e)) \ \textrm{is self-adjoint
  on} \ \H. 
\end{eqnarray}
Let us state three results which will be useful later. 
\begin{lemma} \label{KN-Comp}
  Let $\chi \in \cor$ and $j \in C^\infty(\R)$ a cut-off function at the horizon, i.e. $j = 1$ on a neighbourhood of $-\infty$ and $j = 0$ on a neighbourhood of $+\infty$. Then $j(x) (\chi(H_0) - \chi(H_e))$ is compact on $\H$. 
\end{lemma}
\textbf{Proof}. We use the Helffer-Sj\"ostrand formula (see for instance \cite{Dav}). We obtain
\begin{displaymath}
  j(x) (\chi(H_0) - \chi(H_e)) = \frac{i}{2\pi} \int_\C
  \partial_{\bar{z}} \tilde{\chi}(z) j(x) (z-H_0)^{-1} (H_0 - H_e)
  (z-H_e)^{-1} dz \wedge d\bar{z},
\end{displaymath}
where $\tilde{\chi}$ is an almost analytic extension of $\chi$. The integral converges in operator norm in view of the inequality
$$
\|j(x) (z-H_0)^{-1} (H_0 - H_e)(z-H_e)^{-1} \| \leq C \frac{\kl z \er}{|Im\,z|^2},
$$ 
and thanks to the properties of $\tilde{\chi}$. Hence, it is enough to prove that the operator under the integral is compact. But 
\begin{equation}
\begin{split}
  j(x) (z-H_0)^{-1} (H_0 - H_e)(z-H_e)^{-1} & = (z-H_0)^{-1} j(x) (H_0 - H_e) (z-H_e)^{-1} \\  
                                            & \quad + i(z-H_0)^{-1} j'(x) \Ga (z-H_0)^{-1} (H_0 - H_e)(z-H_e)^{-1},
\end{split}                                             
\end{equation}
Using that $j(x)(H_0 - H_e) = j(x) \big( (a(x) - a_- e^{\kappa_+ x}) \DS + b(x) \Go + (c(x) - c_-) \big)$, both terms are compact by Proposition \ref{PotentialsDO} and the standard compactness criterion Lemma \ref{Compactness}. $\diamondsuit$\\  

\begin{lemma} \label{KN-Smallness-k}
  Let $f, \chi \in C_\infty(\R)$. Then 
  \begin{equation} \label{KN-Small-1}
    f(x + \kappa_+^{-1} \ln|k|) \,\chi(H_e^k) \ \textrm{ is compact on} \ \H_k.
  \end{equation}
  Moreover, for any $\lambda \in \R$ and $\e > 0$, there exists $\delta > 0$ such that  
  \begin{equation} \label{KN-Small-2}
    \| f(x + \kappa_+^{-1} \ln|k|) \mathbf{1}_{[\lambda_0 - \delta, \lambda_0 + \delta]}(H_e^k) \| < \e, \ \textrm{uniformly in} \ k.      
  \end{equation}
\end{lemma}
\textbf{Proof}. For any $k$, the function $g(x) := f(x + \kappa_+^{-1} \ln|k|)$ belongs to $C_\infty(\R)$. Hence the compactness of $f(x + \kappa_+^{-1} \ln|k|) \chi(H_e^{ln})$ follows from (\ref{KN-Comp0}).  

In order to prove (\ref{KN-Small-2}), let us introduce the unitary operator $U^k = e^{-i\kappa_+^{-1} \ln|k| D_x}$. Conjugating the operator in (\ref{KN-Small-2}) by $U^k$, we have to show that for $\delta$ small enough  
\begin{equation} \label{KN-Small-a}
  \|f(x) \mathbf{1}_{[\lambda_0 - \delta, \lambda_0 + \delta]}(\Ga D_x + a_- e^{\kappa_+ x} \Gb + c_-) \| < \e, \ \textrm{uniformly in} \ k. 
\end{equation}
Clearly, the operator in (\ref{KN-Small-a}) is independent of $k$ and is compact by (\ref{KN-Small-1}). Hence the result follows from (\ref{KN-Eigen}). $\diamondsuit$\\
  
This lemma immediately yields the following ``smallness-result'' 
\begin{coro} \label{KN-Smallness}
  Let $f \in C_\infty(\R)$ and $\lambda_0 \in \R$. Then for any $\e >
  0$, there exists $\delta> 0$ such that
  \begin{displaymath}
    \| f(x + \kappa_+^{-1} \ln|\DS|) \mathbf{1}_{[\lambda_0 - \delta,
    \lambda_0 + \delta]}(H_e) \| < \e. 
  \end{displaymath}
\end{coro}

Having recalled these technical results, we turn now our attention towards obtaining a Mourre estimate for $(H_0,A_-)$. 
\begin{lemma} \label{KN-ME-1}
  Let $\lambda_0 \in \R$. Then there exist a function $\chi \in \cor$ with $supp\,\chi$ containing $\lambda_0$, a strictly positive constant $\e$ and a compact operator $K$ on $\H$ such that
  \begin{displaymath}
    \chi(H_0) i[H_0,A_-] \chi(H_0) \geq \e \,\chi(H_0) j^2_1(\frac{x}{S}) \chi(H_0) + K, 
  \end{displaymath}
for $S$ large enough.
\end{lemma}
\textbf{Proof}: Recall that 
\begin{displaymath}
  i[H_0,A_-] = R_-^{(1)}(x,\DS) + 2ia(x) R_-(x,\DS) \DS \Ga +
  2ib(x) R_-(x,\DS) \Go \Ga. 
\end{displaymath}
Let us choose $\chi \in \cor$ such that $supp\,\chi$ contains
$\lambda_0$. We decompose $\chi(H_0) i[H_0,A_-] \chi(H_0)$ into the sum of four
terms  
\begin{displaymath}
  \chi(H_0) i[H_0,A_-] \chi(H_0) = I_1 + I_2 + I_3 + I_4, 
\end{displaymath}
where  
\begin{eqnarray*}
  I_1 & = & \chi(H_0) j_1^2(\frac{x}{S}) \chi(H_0) \\
  I_2 & = & \chi(H_0) j_1^2(\frac{x}{S}) \Big( j_-^2  \big(\frac{x + \kappa_+^{-1} \ln |\DS|}{S} \big) - 1 \Big) \chi(H_0) \\ 
  I_3 & = & \chi(H_0) \Big\{ \frac{x + \kappa_+^{-1} \ln |\DS|}{S} (j_-^2)^{'} \big( \frac{x + \kappa_+^{-1} \ln |\DS|}{S} \big) + 2i a_- e^{(\kappa_+ x + \ln |\DS|)} R_- \frac{\DS}{|\DS|} \Ga \Big\}  \chi(H_0) \\ 
  I_4 & = & \chi(H_0) \Big\{ 2i \frac{a(x) - a_- e^{\kappa_+ x}}{a(x)} R_- a(x) \DS \Ga + 2i b(x) R_- \Go \Ga \Big\} \chi(H_0).      
\end{eqnarray*}
The term $I_4$ is compact on $\H$. For instance, let us treat the first term in $I_4$. We can write it as  
\begin{displaymath}
  - 2i \Big( \chi(H_0) \frac{a(x) - a_- e^{\kappa_+
  x}}{a(x)} j_1^2(\frac{x}{S}) \x \Big) \Big( \x^{-1}
  R_-(x,\DS) \Ga \Big) (a(x) \DS \chi(H_0)).
\end{displaymath}
Using Lemma \ref{PotentialsDO}, we see that the function $\frac{a(x) - a_- e^{\kappa_+ x}}{a(x)} j_1^2(\frac{x}{S}) \x$ vanishes at both $x = \pm \infty$ and thus 
\begin{displaymath} 
  \chi(H_0) \frac{a(x) - a_- e^{\kappa_+ x}}{a(x)} j_1^2(\frac{x}{S}) \x
\end{displaymath}
is compact by Lemma \ref{Compactness}. This implies the compactness of the full term since $\x^{-1} R_-(x,\DS) \Ga$ and $a(x) \DS \chi(H_0)$ are bounded by Lemmas \ref{KN-CO-1} and (\ref{ResolventEstimates}). The other term in $I_4$ is treated similarly. 

Thanks to Lemmata \ref{KN-CO-1} and \ref{KN-Comp}, the term $I_3$ can be
written as
\begin{displaymath}
  \chi(H_e) j_1(\frac{x}{S}) \Big\{ f_S(x + \kappa_+^{-1}
  \ln|\DS|) + g_S (x + \kappa_+^{-1}
  \ln|\DS|) \frac{\DS}{|\DS|} \Ga \Big\} j_1(\frac{x}{S})
  \chi(H_e) + \ K, 
\end{displaymath}
where $K$ compact and $f_S, g_S \in C_\infty(\R)$. By Corollary \ref{KN-Smallness}, the part involving $f_S$ and $g_S$ tends to $0$ in operator norm when $supp\,\chi$ is small enough. More precisely, for all $S$ and $\eo > 0$, we can choose $\chi$ with $supp\,\chi$ small enough such that 
\begin{displaymath}
  I_3 \geq \ -\eo \,\chi(H_0) j_1^2(\frac{x}{S}) \chi(H_0) + \, K.
\end{displaymath}
We now prove that the term $I_2$ is the sum of compact operator plus a term which tends to $0$ in operator norm when $S$ tends to infinity. We first introduce the bounded operator
\begin{displaymath}
  W = j_1(\frac{x}{S}) \eta \big(\frac{x + \kappa_+^{-1} \ln
  |\DS|}{S} \big),  
\end{displaymath}
where $\eta \in C^\infty(\R)$ and $\eta^2 = 1-j_-^2$. Using Lemma \ref{KN-Comp}, we have  
\begin{displaymath}
  I_2 = \chi(H_0) W^2 \chi(H_0) = \chi(H_e) W^2 \chi(H_e) + K,
  \quad K \ \textrm{compact}.  
\end{displaymath}
We claim that 
\begin{equation} \label{KN-ME-1a} 
  \lim_{S \to \infty} \|\chi(H_e) W^2 \chi(H_e)\| = 0.
\end{equation}
To see this, we study the operator $B = \chi(H_e) e^{\kappa_+ x} \DS W^2 e^{\kappa_+ x} \DS \chi(H_e)$. Since $W$ preserves $D(H_e)$ (\cite{HaN}, Lemma 5.17), $B$ is well defined on $\H$. Moreover, using (\ref{KN-ResEst-4}), $B$ is in fact bounded. Now, since $W$ commutes with $e^{\kappa_+ x} \DS$, $B$ can be written as $\chi(H_e) W e^{2  \kappa_+ x} \DS^2 W \chi(H_e)$. But, note that on $supp\, \eta$, we have  
\begin{displaymath}
  e^{\kappa_+ x} |\DS| \geq \ e^{\kappa_+ S}.
\end{displaymath}
Hence, $B \geq e^{2\kappa_+ S} \chi(H_e) W^2 \chi(H_e)$ and since $B$ is bounded, there exists a constant $C$ such that 
\begin{displaymath}
  \chi(H_e) W^2 \chi(H_e) \leq C e^{-2\kappa_+ S} \ \to \ 0, \quad S
  \to \infty,
\end{displaymath}
which implies (\ref{KN-ME-1a}). Finally, using that $[\chi(H_e), j_1(\frac{x}{S})] \in O(S^{-1})$ and the previous estimate, we obtain: for any $\eo > 0$, we can find $S_0$ large enough such that for any $S > S_0$,  
\begin{displaymath} 
  \chi(H_0) W^2 \chi(H_0) \geq \ -\eo \, \chi(H_0) j_1^2(\frac{x}{S})
  \chi(H_0) + \, K. 
\end{displaymath}
This concludes the proof of the lemma. $\diamondsuit$\\  

\noindent \textbf{The Mourre estimate for $(H_0,A_+)$}: The situation at infinity is more standard. The only subtlety comes from the choice of a conjugate operator since this choice depends on the energy interval we consider. In particular, two threshold values appear for which we are not able to establish a Mourre estimate. Precisely, we have
\begin{lemma} \label{KN-ME-2} 
  \begin{itemize} 
    \item [(a)] For any $\chi \in \cor$ with $supp\,\chi \subset (+m, +\infty)$, there exist $\e > 0$ and a compact operator $K$ such that 
    \begin{equation} \label{KN-ME-21}
      \chi(H_0) i[H_0,A_+] \chi(H_0) \geq \ \e \, \chi(H_0)
      j_+^2(\frac{x}{S}) \chi(H_0) + K. 
    \end{equation}
    \item [(b)] For any $\chi \in \cor$ with $supp\,\chi \subset (-\infty, -m)$, (\ref{KN-ME-21}) is true if we replace $A_+$ by $-A_+$.  
    \item [(c)] For any $\chi \in \cor$ with $supp\,\chi \subset  (-m,+m)$, the operator $j^2_+(\frac{x}{S}) \chi(H_0)$ is compact on  $\H$ and (\ref{KN-ME-21}) is valid for any $\e > 0$.  
  \end{itemize}  
\end{lemma}
\textbf{Proof}: Let us show (a). Using (\ref{ResolventEstimates}) and the fact that $j_+^{'}(\frac{x}{S}) \chi(H_0)$ is compact, we obtain
\begin{eqnarray*}
  \chi(H_0) i[H_0,A_+] \chi(H_0) & = & \chi(H_0) j_+(\frac{x}{S}) \Ga  D_x j_+(\frac{x}{S}) \chi(H_0) \\
  & & - \chi(H_0) R_+(x) \Big\{ a^{'}(x) \DS + b^{'}(x) \Go  + c^{'}(x) \Big\} \chi(H_0) + K, 
\end{eqnarray*}
where $K$ compact. We now make $H_0$ appear in the first term and using that the three operators
\begin{eqnarray*}
  \chi(H_0) j_+^2(\frac{x}{S}) \Big\{ x a^{'}(x) + a(x)
  \Big\} \DS \chi(H_0), \\
  \chi(H_0) \Big\{ R_+(x) b^{'}(x) + (b(x) - m) \Big\} \Go
  \chi(H_0), \\
  \chi(H_0) \Big\{ R_+(x) c^{'}(x) + c(x) \Big\} \chi(H_0), 
\end{eqnarray*}
are compact by Proposition \ref{PotentialsDO} and Lemma \ref{Compactness}, we get
\begin{displaymath}
  \chi(H_0) i[H_0,A_+] \chi(H_0) = \chi(H_0) j_+(\frac{x}{S}) H_0
  j_+(\frac{x}{S}) \chi(H_0) - \chi(H_0) j_+^2(\frac{x}{S}) m 
  \Go \chi(H_0) + K.
\end{displaymath}
Finally, using again that $j_+^{'}(\frac{x}{S}) \chi(H_0)$ is compact and the fact that $supp\,\chi \subset (+m,+\infty)$, there exists a strictly positive constant $\e$ such that 
\begin{displaymath}
  \chi(H_0) j_+(\frac{x}{S}) H_0 j_+(\frac{x}{S}) \chi(H_0)  \geq (m + \e) \, \chi(H_0) j_+^2(\frac{x}{S}) \chi(H_0) + K, 
\end{displaymath}
while the second term is obviously estimated by
\begin{displaymath}
  \chi(H_0) j_+^2(\frac{x}{S}) m \Go \chi(H_0) \geq -m \, \chi(H_0) j_+^2(\frac{x}{S}) \chi(H_0). 
\end{displaymath}
This implies (\ref{KN-ME-21}). 

The proof of (b) is identical to the preceding one with $A_+$ replaced by $-A_+$. 

In order to prove (c), we introduce the self-adjoint operator $H_m = \Ga D_\rho + \frac{1}{\rho} \DS + m \Go$ acting on $\H_\infty = L^2(\R^+_\rho \times S^3; \rho^2 d\rho d\omega, \C^4)$. The operator  $H_m$ is exactly the free Dirac Hamiltonian in 5D-flat space-time written in polar coordinates. Observe that $H_m$ corresponds to the formal limit of $H_0$ when $x \to + \infty$ with the identification $\rho = x$. It is well known that (see for instance \cite{Th}) 
\begin{displaymath}
  \sigma(H_m) =\sigma_{ac}(H_m) = (-\infty,-m] \cup [+m,+\infty),
\end{displaymath}
which implies  
\begin{equation} \label{KN-ChiHm} 
  \chi(H_m) = 0,
\end{equation}
if $supp\,\chi \subset (-m,+m)$. Using (\ref{KN-ChiHm}), we can express $\chi(H_0)$ as the difference $\chi(H_0) - \chi(H_m)$ and we want to use the Helffer-Sj\"ostrand formula and the standard compactness criterion to prove (c). Since $H_0$ and $H_m$ do not act on the same Hilbert space, we have to be cautious with this procedure. We proceed as follows. Since $[j_+(\frac{x}{S}), \chi(H_0)]$ is compact, it is enough to prove that $j_+(\frac{x}{S}) \chi(H_0) j_+(\frac{x}{S})$ is compact. Remark that the cut-off function $j_+(\frac{x}{S})$ obviously plays the role of a (bounded) identification operator between $\H$ and $\H_\infty$ (and conversely). Indeed, we have \begin{eqnarray*}
  \mathcal{I}: \ \ \quad  \H \ \longrightarrow \ \H_\infty, \quad \quad
  \quad  & \quad \quad \mathcal{I}^*: &   
  \H_\infty \longrightarrow \H, \\
               \psi(x) \longrightarrow j_+(\frac{\rho}{S})
  \psi(\rho),  & & \psi(\rho) \longrightarrow j_+(\frac{x}{S})
  \psi(x), \ x \geq 0 \ \ \textrm{and} \ 0 \ \textrm{otherwise}. 
\end{eqnarray*}
Hence, using these identification operators and (\ref{KN-ChiHm}), the following identity makes sense on $\H$ 
\begin{displaymath}
  j_+(\frac{x}{S}) \chi(H_0) j_+(\frac{x}{S}) = j_+(\frac{x}{S})
  \big( \chi(H_0) - \chi(H_m) \big) j_+(\frac{x}{S}).
\end{displaymath}
Now using the Helffer-Sj\"ostrand formula, it suffices to show that 
\begin{displaymath}
  L = j_+(\frac{x}{S}) \big( (z-H_0)^{-1} - (z-H_m)^{-1} \big)
  j_+(\frac{x}{S}),   
\end{displaymath}
is compact on $\H$. We introduce $\tilde{j}_+ \in C^\infty(\R)$
satisfying $\tilde{j}_+ j_+ = j_+$ and $\tilde{j}_+ = 0$ on
$(-\infty,0]$. Then we have 
\begin{eqnarray*}
  L & = & j_+(\frac{x}{S}) \big( (z-H_0)^{-1}
  \tilde{j}_+(\frac{x}{S}) - \tilde{j}_+(\frac{x}{S}) (z-H_m)^{-1}
  \big) j_+(\frac{x}{S}), \\
    & = & j_+(\frac{x}{S}) (z-H_0)^{-1} \big( H_0
  \tilde{j}_+(\frac{x}{S}) - \tilde{j}_+(\frac{x}{S}) H_m \big)
  (z-H_m)^{-1} j_+(\frac{x}{S}), \\
    & = & j_+(\frac{x}{S}) (z-H_0)^{-1} \Big\{ -\frac{i}{S} \Ga
  (\tilde{j}_+)^{'}(\frac{x}{S}) + \tilde{j}_+(\frac{x}{S})
  (a(x) - \frac{1}{x}) \DS \\ & & \hspace{3cm} +
  \tilde{j}_+(\frac{x}{S}) 
  (b(x) - m) \Go + \tilde{j}_+(\frac{x}{S}) c(x) \Big\}
  (z-H_m)^{-1} j_+(\frac{x}{S}). 
\end{eqnarray*}
Thus $L$ is compact using Proposition \ref{PotentialsDO} and Lemma \ref{Compactness}. This concludes the proof of the lemma. $\diamondsuit$\\  

Summarizing all the previous results, we have established a Mourre estimate for the pair of Hamiltonians $(H_0,A_\chi)$. Precisely, we have proved
\begin{prop} \label{KN-ME-3}
  Let $\chi \in \cor$ such that $supp\,\chi$ is included in one of the intervals $(-\infty,-m)$, $(-m,+m)$, $(+m,+\infty)$ and $supp\,\chi$ small enough. Then for $S$ sufficiently large, there exist $\e > 0$ and a compact operator $K$ such that
  \begin{displaymath}
    \chi(H_0) i[H_0,A_\chi] \chi(H_0) \geq \ \mu \, \chi^2(H_0) + K, 
  \end{displaymath}
  where $A_\chi = A_- + A_+$ when $supp\,\chi \subset (+m,+\infty)$, $A_\chi = A_- - A_+$ when $supp\,\chi \subset (-\infty,-m)$ and $A_\chi = A_-$ when $supp\,\chi \subset (-m,+m)$. 
\end{prop}

\noindent \textbf{The Mourre estimate for $(H,A_\chi)$}: Let $\chi \in \cor$ satisfying the assumptions of Proposition \ref{KN-ME-3}. We only treat the case $supp\,\chi \subset (+m, +\infty)$ since the other cases are analogous. We have $A_\chi = A = A_- + A_+$, and using Lemma \ref{KN-CO-4} and Corollary \ref{CompactnessH-HO}, we have   
\begin{displaymath}
  \chi(H) i[H,A] \chi(H) = \chi(H_0) i[H,A] \chi(H_0) + K, 
\end{displaymath}
where $K$ compact. We write $H = H_0 + (h-1) H_0 + H_0 (h-1) + (h-1)
H_0 (h-1) + M = H_0 + H_1$ and we show that
\begin{displaymath}
  \chi(H_0) i[H_1,A] \chi(H_0) \ \ \textrm{is compact}. 
\end{displaymath}
For instance, we have
\begin{displaymath}
  \chi(H_0) i[(h-1) H_0,A] \chi(H_0) = \chi(H_0) i[(h-1),A]
  H_0 \chi(H_0) + \chi(H_0) (h-1) i[H_0,A] \chi(H_0).
\end{displaymath}
But, $i[(h-1), A] = -R_+(x) \partial_{x} h$ and $h-1$ belong to $S^{-2}$ by Lemma \ref{PotentialsD}. Hence this term is compact. The other terms involving $h-1$ are treated similarly. Finally, since $M \in S^{-2}$ according to Lemma \ref{PotentialsD}, it is immediate that $[M,A] \in S^{-1}$ and thus, the last term $\chi(H_0) i[M,A] \chi(H_0)$ is also compact. Hence, we have proved 
\begin{prop} \label{KN-ME-4}
  Let $\chi \in \cor$ such that $supp\,\chi$ is included in one of the intervals $(-\infty,-m)$, $(-m,+m)$, $(+m,+\infty)$ and $supp\,\chi$ small enough. Then for $S$ sufficiently large, there exist $\e > 0$ and a compact operator $K$ such that
  \begin{displaymath}
    \chi(H) i[H,A_\chi] \chi(H) \geq \ \mu \, \chi^2(H) + K, 
  \end{displaymath}
  where $A_\chi$ defined as in Proposition \ref{KN-ME-3}. 
\end{prop}

We finish this section with some direct corollaries of Theorem \ref{KN-MO} and Propositions \ref{KN-ME-3} and \ref{KN-ME-4}. The first result concerns the spectrum of $H_0$. Using the fact that $H_0$ can be decomposed into a family of one-dimensional Dirac operators (see Lemma \ref{SphericalHarmonics}), we can readily prove that the spectrum of $H_0$ is continuous (no pure point spectrum) by the same argument as in Proposition \ref{NoPurePointSpectrum}. Together with the Mourre estimate obtained in Proposition \ref{KN-ME-3}, we thus get  
\begin{thm} \label{KN-Spectrum-H0} 
  $\sigma(H_0) = \sigma_{ess}(H_0) = \sigma_{ac}(H_0) = \R$. 
\end{thm}


Since $(H-i)^{-1} - (H_0-i)^{-1}$ is compact (see the proof of Corollary \ref{CompactnessH-HO}), it follows from the Weyl Theorem (see \cite{RS}, Vol IV) and Theorem \ref{KN-Spectrum-H0} that $\sigma_{ess}(H) = \R$. Moreover, applying Proposition \ref{KN-ME-4} and Theorem \ref{KN-MO} again, we obtain the following result for the spectrum of $H$
\begin{thm} \label{KN-Spectrum-H}
  The operator $H$ has no singular continuous spectrum ($\sigma_{sing}(H) = \emptyset$), $\sigma_{ess}({H}) = \R$ and $H$ has only
  eigenvalues of finite multiplicity in $\R \setminus\{\pm m\}$.     
\end{thm}
Note here that we cannot say that $H$ has no eigenvalue a priori since $H$ is different from $\DR$ by definition. However, since $H$ coincides with $\DR$ on the fixed angular mode $e^{in\varphi} e^{im\psi}$ and since $\sigma(\DR) = \emptyset$, then $\sigma_{pp}(H^{nm}) = \emptyset$

\subsection{Local energy decay II}

We prove now our main result that can be restated (in the variable $x$) as follows
\begin{thm}
  The Dirac operator $\DR$ is a self-adjoint operator on $\H$ having purely absolutely continuous spectrum. Moreover, for all $\chi \in \cor$ and all $u \in \H$, we have the following local energy decay
\begin{equation} \label{Decay}
  \lim_{t \to \pm \infty} \| \chi(x) e^{-it\DR} u \| = 0.
\end{equation}
\end{thm}
\textbf{Proof}. Since $\DR$ coincides with $H$ on the angular mode $e^{in\varphi} e^{im\psi}$, we conclude from Theorem \ref{KN-Spectrum-H} that $\sigma_{\textrm{sc}}(\DR^{nm}) = \emptyset$. Since this can be done for all $n,m \in \frac{1}{2} + \Z$, the first assertion follows. 

By a density argument, it is enough to prove (\ref{Decay}) on each angular mode $\{e^{in\varphi} e^{im\psi}\}$, $n,m \in \frac{1}{2} + \Z$ and for all $u$ in a domain dense in $\H$, for instance $D(\DR)$. But since the spectrum of $\DR$ is absolutely continuous and since $\chi(x) (\DR^{nm} + i)^{-1}$ is compact on $\H_{nm}$ by Lemma \ref{Compactness}, we have for all $u_{nm} \in D(\DR^{nm})$
$$
  \lim_{t \to \pm \infty} \| \chi(x) e^{-it\DR^{nm}} u_{nm} \| = 0,
$$
by the Riemann-Lebesgue Lemma (see \cite{RS}). This finishes the proof of our main result. $\diamondsuit$ \\


\appendix
\Section{The Dirac equation adapted to the separation and absence of pure point spectrum} \label{Separation}

In this Appendix, we first recall the expression of the Dirac equation obtained by Wu \cite{Wu} and follow his argument to show the separability of the equation into radial and angular systems of ODEs. Using this form of the equation, we are able to prove that the Dirac equation has no eigenmode, or equivalently that the pure point spectrum of the corresponding Hamiltonian is empty. In the course of the proof, we recall how to decompose the angular operator resulting from the separation of variables in a well chosen Hilbert basis of eigenfunctions. This is also the result we apply to decompose the standard Dirac operator $\DS$ on $S^3$ in Sections \ref{SpectralProperties} and \ref{Mourre}. 

Using the local Lorentz frame for the metric (\ref{Metric}) given by
\begin{equation}
  \begin{split}
    f_0 & = \frac{(r^2+a^2)(r^2+b^2)}{r \sqrt{\Delta \Sigma}} \Big( \partial_t + \frac{a}{r^2+a^2} \,\partial_\varphi + \frac{b}{r^2+b^2} \,\partial_\psi \Big), \\
    f_1 & = \sqrt{\frac{\Delta}{r^2 \Sigma}} \, \partial_r, \\
    f_2 & = \frac{1}{\sqrt{\Sigma}} \, \partial_\theta, \\
    f_3 & = \frac{\sin\theta \cos\theta}{p \sqrt{\Sigma}} \Big( (a^2-b^2) \, \partial_t + \frac{a}{\sin^2\theta} \, \partial_\varphi - \frac{b}{\cos^2\theta} \, \partial_\psi \Big), \\
    f_5 & = \frac{1}{rp} \Big( ab \, \partial_t + b \, \partial_\varphi + a \, \partial_\psi \Big), 
  \end{split}
\end{equation}
where $p^2 = a^2 \cos^2\theta + b^2 \sin^2\theta$, Wu obtained the following expression for the Dirac equation (\ref{DE})
\begin{equation}
  \begin{split}
    & \Big[ \gamma^0 \frac{(r^2+a^2)(r^2+b^2)}{r\sqrt{\Delta \Sigma}} \Big( \partial_t + \frac{a}{r^2+a^2} \, \partial_\varphi + \frac{b}{r^2+b^2} \, \partial_\psi \Big) + \gamma^1 \sqrt{\frac{\Delta}{r^2\Sigma}} \Big( \partial_r + \frac{\partial_r\Delta}{4\Delta} + \frac{r-ip\gamma^5}{2\Sigma} \Big) \\
    & \hspace{1cm} + \gamma^2 \frac{1}{\sqrt{\Sigma}} \Big( \partial_\theta + \frac{\cot\theta}{2} - \frac{\tan \theta}{2} - \frac{(a^2-b^2) \sin\theta \cos\theta}{2\Sigma p} i \gamma^5 (r-ip\gamma^5) \Big) \\
    & \hspace{2cm} + \gamma^3 \frac{\sin\theta \cos\theta}{p \sqrt{\Sigma}} \Big( (a^2-b^2) \, \partial_t + \frac{a}{\sin^2\theta} \, \partial_\varphi - \frac{b}{\cos^2\theta} \, \partial_\psi \Big) \\
    & \hspace{3cm} + \gamma^5 \frac{1}{rp} \Big( ab \partial_t + b \partial_\varphi + a \partial_\psi \Big) + \frac{i ab}{2r^2p^2} \gamma^0 \gamma^1 (r+ip\gamma^5) + m \Big] \phi = 0. 
  \end{split}
\end{equation}
Let us choose a square root of $r+ip\gamma^5$ - for instance $\sqrt{r+ip\gamma^5} = \sqrt{\frac{r + \sqrt{\Sigma}}{2}} \, I_4 + i \sqrt{\frac{\sqrt{\Sigma}-r}{2}} \, \gamma^5$ - and introduce the weighted spinor $v = \Delta^{\frac{1}{4}} \sqrt{r+ip\gamma^5} \phi$. Then $v$ satisfies 
\begin{equation} \label{DES-1}
  \begin{split}
    & \Big[ \Big( \gamma^0 \frac{(r^2+a^2)(r^2+b^2)}{r\sqrt{\Delta \Sigma}} + \gamma^3 \frac{\sin\theta \cos\theta(a^2-b^2)}{p} + \big(\frac{\gamma^5}{p} - \frac{i}{r} \big) ab \Big) \partial_t + \gamma^1 \sqrt{\frac{\Delta}{r^2}} \partial_r + \\
    & \hspace{0.5cm} + \gamma^2 \Big( \partial_\theta + \frac{\cot\theta}{2} - \frac{\tan \theta}{2} + \gamma^3 \frac{1}{p} \Big( a\cot\theta \partial_\varphi - b\tan\theta\partial_\psi \Big) + \gamma^5 \frac{1}{p} \Big( b \partial_\varphi + a \partial_\psi \Big) \\
    & \hspace{1cm} + \Big( \gamma^0 \frac{a(r^2+b^2)}{r\sqrt{\Delta}} - \frac{ib}{r} \Big) \partial_\varphi + \Big( \gamma^0 \frac{b(r^2+a^2)}{r\sqrt{\Delta}} - \frac{ia}{r} \Big) \partial_\psi + \frac{i ab}{r^2} \gamma^0 \gamma^1 + mr - imp\gamma^5 \Big] v = 0. 
  \end{split}
\end{equation}
The matrix-valued function $P = \Big( \gamma^0 \frac{(r^2+a^2)(r^2+b^2)}{r\sqrt{\Delta \Sigma}} + \gamma^3 \frac{\sin\theta \cos\theta(a^2-b^2)}{p} + \gamma^5 \frac{ab}{p} - \frac{iab}{r} \Big)$ in front of $\partial_t$ is invertible with inverse
$$
  P^{-1} = -\frac{\Delta}{\tau} \Big( \gamma^0 \frac{(r^2+a^2)(r^2+b^2)}{r\sqrt{\Delta \Sigma}} + \gamma^3 \frac{\sin\theta \cos\theta(a^2-b^2)}{p} + \frac{ab}{p} \gamma^5 + \frac{iab}{r} \Big). 
$$
Multiplying (\ref{DES-1}) by $P^{-1}$ from the left and introducing the Regge-Wheeler variable $x$ as in (\ref{RW}), we thus obtain
\begin{equation} \label{DES-2}
  \begin{split}
    i \partial_t v & = P^{-1} \Big[ \gamma^1 \frac{(r^2+a^2)(r^2+b^2)}{r\sqrt{\Delta}} D_x + \DS + \Big( \gamma^0 \frac{a(r^2+b^2)}{r\sqrt{\Delta}} - \frac{ib}{r} \Big) D_\varphi + \Big( \gamma^0 \frac{b(r^2+a^2)}{r\sqrt{\Delta}} - \frac{ia}{r} \Big) D_\psi \\
    & \hspace{3cm} + \frac{ab}{r^2} \gamma^0 \gamma^1 - i mr - mp\gamma^5 \Big] v.
  \end{split}  
\end{equation}
Using the notations introduced in (\ref{Notations}), we can write $P^{-1}$ as
$$
  P^{-1} = - N \frac{r\sqrt{\Delta}}{(r^2+a^2)(r^2+b^2)} \gamma^0,
$$
where
\begin{equation} \label{N}
  N = h^4 \Big[ I_4 - \frac{r\sqrt{\Delta}}{(r^2+a^2)(r^2+b^2)} \Big( \frac{(a^2-b^2)}{p} \sin\theta \cos\theta \Gc + \frac{ab}{p} \Gd + \frac{ab}{r} \Go \Big) \Big], 
\end{equation}
and we get for the Dirac equation (\ref{DES-2})
\begin{equation} \label{DES-3}
  \begin{split}
    i \partial_t v & = N \Big[ \Ga D_x + \frac{r\sqrt{\Delta}}{(r^2+a^2)(r^2+b^2)} \DS + \Big( \frac{a}{r^2+a^2} + \frac{b\sqrt{\Delta}}{(r^2+a^2)(r^2+b^2)} \Go \Big) D_\varphi \\
    & \hspace{1cm} + \Big( \frac{b}{r^2+b^2} + \frac{a\sqrt{\Delta}}{(r^2+a^2)(r^2+b^2)} \Go \Big) D_\psi + \frac{ab\sqrt{\Delta}}{r(r^2+a^2)(r^2+b^2)} \gamma^1 \\
    & \hspace{2cm} + \frac{m r^2\sqrt{\Delta}}{(r^2+a^2)(r^2+b^2)} \Go - \frac{mp r\sqrt{\Delta}}{(r^2+a^2)(r^2+b^2)} \Gd \Big] v.
  \end{split}  
\end{equation}
Let us denote 
\begin{equation} \label{DOS}
  \D_0 = \DO + \frac{\sqrt{\Delta}}{(r^2+a^2)(r^2+b^2)}\Big( b\Go D_\varphi + a \Go D_\psi + \frac{ab}{r} \gamma^1 - mpr\Gd \Big).
\end{equation}
We finally get the following Hamiltonian form for (\ref{DES-3})
\begin{equation} \label{DES}
\begin{split}
  & i \partial_t v = \D v, \\
  & \D = N \D_0,
\end{split}
\end{equation}
with $N$ and $\D_0$ given by (\ref{N}) and (\ref{DOS}) respectively. 

The Hilbert space framework for the Hamiltonian $\D$ is as follows. We define the Hilbert space 
$$
  \G = L^2(\R \times S^3, dx d\omega; \C^4),
$$
equipped with the scalar product
\begin{equation} \label{ScalarProduct}
  ( u,v )_\G = \kl N^{-1} u,v \er_\H,
\end{equation}
where $N^{-1}$ is shown to be
$$
  N^{-1} = \Big[ I_4 + \frac{r\sqrt{\Delta}}{(r^2+a^2)(r^2+b^2)} \Big( \frac{(a^2-b^2)}{p} \sin\theta \cos\theta \Gc + \frac{ab}{p} \Gd + \frac{ab}{r} \Go \Big) \Big]. 
$$
Note that the symmetric bilinear form (\ref{ScalarProduct}) is indeed a scalar product by the following argument. Denote $N^{-1} = (I_4 + M)$. Then a short calculation shows that $(I_4 + M)(I_4 - M) = I_4 - M^2 = h^{-4}$. Hence  
$$
  M^2 = \frac{h^4 - 1}{h^4} = \frac{\Delta ( r^2 p^2 + a^2 b^2)}{(r^2+a^2)^2(r^2+b^2)^2}.
$$
Since $M^* = M$, we conclude that
$$
  \| M\|^2 = \frac{\Delta ( r^2 p^2 + a^2 b^2)}{(r^2+a^2)^2(r^2+b^2)^2} < \frac{r^2 (a^2+b^2) + a^2 b^2}{(r^2+a^2)(r^2+b^2)} < 1.
$$
As a consequence
$$
  N^{-1} = (I + M) \geq 0.  
$$

Finally, according to (\ref{DOS}), the Hamiltonian $\D_0$ is clearly a short-range perturbation of the Hamiltonian $\DO$ when restricted to each angular mode $e^{in\varphi} e^{im\psi}$. By the same argument as in Section \ref{SpectralProperties}, $\D_0$ is thus self-adjoint on $\H$ with its natural domain $D(\D_0) = \{ v \in \H, \ \|\D_0 v \|^2  < \infty \}$. Moreover, since the operator $N$ is clearly a bounded self-adjoint operator on $\H$, we have
\begin{lemma}
  The Hamiltonian $\D$ is self-adjoint on $\G$ with the same domain as $\D_0$, \textit{i.e.} $D(\D) = D(\D_0)$. 
\end{lemma}

We now prove the absence of pure point spectrum for $\D$. In the course of the proof, we shall see how the form of the equation (\ref{DES}) allows easily to separate the equations into radial and angular systems of ODEs which, in turn, permit us to prove the result. 

\begin{prop} \label{NoPurePointSpectrum}
  $\sigma_{\textrm{pp}}(\D) = \emptyset$. 
\end{prop}
\textbf{Proof}. Suppose that $\sigma_{\textrm{pp}}(\D) \ne \emptyset$. There exist thus $\omega \in \R$ and $v \in \G$, $v \ne 0,$ such that $\D v = \omega v$. Then $N(\D_0 - \omega N^{-1}) \, v = 0$. More explicitly, we have
\begin{equation} \label{S-1}
\begin{split}
  & \Big\{ \Big[ \Ga D_x + \frac{r\sqrt{\Delta}}{(r^2+a^2)(r^2+b^2)} \DS + \Big( \frac{a}{r^2+a^2} + \frac{b\sqrt{\Delta}}{(r^2+a^2)(r^2+b^2)} \Go \Big) D_\varphi \\
  & + \Big( \frac{b}{r^2+b^2} + \frac{a\sqrt{\Delta}}{(r^2+a^2)(r^2+b^2)} \Go \Big) D_\psi + \frac{ab\sqrt{\Delta}}{r(r^2+a^2)(r^2+b^2)} \gamma^1 \\
  & \hspace{1cm} + \frac{m r^2\sqrt{\Delta}}{(r^2+a^2)(r^2+b^2)} \Go - \frac{mp r\sqrt{\Delta}}{(r^2+a^2)(r^2+b^2)} \Gd \Big] \\
  & \hspace{2cm} - \omega \Big[ I_4 + \frac{r\sqrt{\Delta}}{(r^2+a^2)(r^2+b^2)} \Big( \frac{(a^2-b^2)}{p} \sin\theta \cos\theta \Gc + \frac{ab}{p} \Gd + \frac{ab}{r} \Go \Big) \Big] \Big\}= 0.
\end{split} 
\end{equation}
The above equation (\ref{S-1}) can be rewritten as 
\begin{equation}
  \frac{r\sqrt{\Delta}}{(r^2+a^2)(r^2+b^2)} \Big[ \Ra + \A \Big] v = 0,
\end{equation}
where
\begin{equation}
\begin{split}
  \Ra & = \frac{(r^2+a^2)(r^2+b^2)}{r\sqrt{\Delta}} \Ga D_x + mr\Go + \Big( \frac{a(r^2+b^2)}{r\sqrt{\Delta}} + \frac{a}{r} \Go \Big) D_\varphi \\
      & \hspace{1cm} + \Big( \frac{b(r^2+a^2)}{r\sqrt{\Delta}} + \frac{b}{r} \Go \Big) D_\psi + \frac{ab}{r^2} \gamma^1 - \omega \Big( \frac{(r^2+a^2)(r^2+b^2)}{r\sqrt{\Delta}} + \frac{ab}{r} \Go \Big), 
\end{split}
\end{equation}
and
\begin{equation}      
  \A  = \DS - m p \Gd - \omega \Big( \frac{(a^2-b^2) \sin\theta \cos \theta}{p} \Gc + \frac{ab}{p} \Gd \Big).        
\end{equation}
Note that, once restricted to the angular modes $\{e^{in\varphi} e^{im\psi}\}$, $n,m \in \frac{1}{2} + \Z$, the operators $\R^{nm}$ and $\A^{nm}$ are two matrix ordinary differential operators in the $r$ and $\theta$ variables respectively, hence the separation of variables. We shall use this property as follows
\begin{lemma} \label{SeparationA}
  The Hilbert space $\H$ can be decomposed onto a Hilbert sum 
  $$
    \H = \displaystyle\bigoplus_{l,n,m \in \mathcal{L}} \H_{lnm}, \quad \mathcal{L} = \big\{(l,n,m), \ \ l \in \N^*, \ n,m \in \frac{1}{2} + \Z \big\},
  $$
  where the $\H_{lnm}$'s are subpaces of $\H$, isometric to $L^2(\R,dx; \C^4)$, and invariant with respect to the actions of $\A, D_\varphi, D_\psi$ and $\Ra$. Precisely, for all $v_{lnm} \in \H_{lnm} \simeq L^2(\R,dx;\C^4)$, the following properties hold: \\
  
  1) $\A v_{lnm} := \A^{lnm} v_{lnm} = \lambda_{lnm}(\omega) \,\Gb v_{lnm}$, \\
  
  2) $\D_\varphi v_{lnm} = n \,v_{lnm}, \ D_{\psi} v_{lnm} = m \,v_{lnm}$, \\ 
  
  3) $\Ra v_{lnm} := \Ra^{lnm} v_{lnm}$ where
  \begin{equation}
  \begin{split}
    \Ra^{lnm} & = \frac{(r^2+a^2)(r^2+b^2)}{r\sqrt{\Delta}} \Ga D_x + mr\Go + \Big( \frac{an(r^2+b^2)}{r\sqrt{\Delta}} + \frac{an}{r} \Go \Big)  + \Big( \frac{bm(r^2+a^2)}{r\sqrt{\Delta}} + \frac{bm}{r} \Go \Big) \\
              & \hspace{2cm} + \frac{ab}{r^2} \gamma^1 - \omega \Big( \frac{(r^2+a^2)(r^2+b^2)}{r\sqrt{\Delta}} + \frac{ab}{r} \Go \Big).
  \end{split}            
  \end{equation}
\end{lemma}

Assuming for the moment this Lemma to be true, we see that the components $v_{lnm} \in L^2(\R,dx;\C^4)$ of the eigenfunction $v$ of $\D$ satisfy the system of equations $\Ra^{lnm} v_{lnm} + \lambda_{lnm} \Gb v_{lnm} = 0$ which can be equivalently written as
\begin{equation} \label{S-2}
\begin{split}
  & \Big[ \partial_x + i \Big(\frac{an}{(r^2+a^2)} + \frac{bm}{(r^2+b^2)} - \omega \Big) \Ga \Big] v_{lnm} = \\
  & \hspace{2cm} -i \frac{r\sqrt{\Delta}}{(r^2+a^2)(r^2+b^2)} \Ga \Big[ mr \Go + \frac{an + bm}{r} \Go \frac{ab}{r^2} \gamma^1 - \omega \frac{ab}{r} \Go + \lambda_{lnm}(\omega) \Gb \Big] v_{lnm}. 
\end{split}
\end{equation}
Let us abbreviate (\ref{S-2}) as
\begin{equation} \label{S-3}
  [\partial_x + iP_{nm}(x,\omega) \Ga ] v_{lnm} = V_{lnm}(x,\omega) v_{lnm},
\end{equation}   
where we let
\begin{equation} \label{Pnm}
  P_{nm}(x,\omega) := \Big(\frac{an}{(r^2+a^2)} + \frac{bm}{(r^2+b^2)} - \omega \Big),
\end{equation}
and 
\begin{equation} \label{Vnm}
  V_{lnm}(x,\omega): = -i \frac{r\sqrt{\Delta}}{(r^2+a^2)(r^2+b^2)} \Ga \Big[ mr \Go + \frac{an + bm}{r} \Go \frac{ab}{r^2} \gamma^1 - \omega \frac{ab}{r} \Go + \lambda_{lnm}(\omega) \Gb \Big].
\end{equation}
Remark that $V_{lnm}(x,\omega) \in L^1(\R^-,dx \,;M_4(\C))$ thanks to the exponential decay of $\Delta$ at the event horizon. We now set $w_{lnm} = \textrm{exp}(iP_{nm}(x,\omega) \Ga) v_{lnm}$. Then the components $w_{lnm}$ clearly belong to $L^2(\R,dx;\C^4)$ and satisfy 
\begin{equation} \label{S-4}
  \partial_x w_{lnm} = \textrm{exp}(iP_{nm}(x,\omega) \Ga) V_{lnm}(x,\omega) \textrm{exp}(-i P_{nm}(x,\omega) \Ga) w_{lnm} = W_{lnm}(x,\omega) w_{lnm}.
\end{equation}
Moreover, from (\ref{Pnm}) and (\ref{Vnm}), we have
$$
  W_{lnm}(x,\omega) \in C^\infty(\R; \,M_4(\C)) \cap L^1(\R^-, dx; \, M_4(\C)).
$$
Since $W_{lnm}$ belong to $L^1$ near the event horizon, by a standard argument, there exists a unique solution of (\ref{S-4}) satisfying $w_{lnm}(-\infty) = C$ for all $C \in \C^4$. The set of solutions of (\ref{S-4}) being of complex dimension $4$, this property characterizes all possible solutions of (\ref{S-4}). As a consequence, we conclude that the unique solution of (\ref{S-4}) belonging to $L^2(\R,dx;\C^4)$ is $0$ which immediately leads to a contradiction. Hence $\sigma_\textrm{pp}(\D) = \emptyset$. $\diamondsuit$ \\

\noindent\textbf{Proof of Lemma \ref{SeparationA}}. We prove this Lemma in two steps. First, using the fact that the operator $\A$ is a bounded perturbation of $\DS$, we show that $\A$ has compact resolvent. There exists therefore a sequence of vector-valued functions $Y_{lnm}$ which are eigenfunctions of $\A$ (and also of $D_\varphi, D_\psi$) that forms a Hilbert basis of $L^2(S^3, d\omega, \C^4)$. Second, we construct explicitely the subspaces $\H_{lnm}$ with the required properties in terms of the eigenfunctions $Y_{lnm}$. 

Recall first that $\DS$ is the standard Dirac operator on $S^3$ associated to the metric $g_{S^3} = d\theta^2 + \sin^2 \theta d\varphi^2 + \cos^2 \theta d\psi^2$ with the ranges $0 \leq \theta \leq \frac{\pi}{2}$ and $0 \leq \varphi, \psi \leq 2\pi$. The operator $\DS$ is self-adjoint on $L^2(S^3, d\omega, \C^4)$ with its natural domain and it is well known that it has purely discrete spectrum (see for instance \cite{CH}) and has compact resolvent. Since the operator $\A$ is clearly a bounded symmetric perturbation of $\DS$ for any $\omega \in \R$, we conclude that $\A$ is also self-adjoint on $L^2(S^3, d\omega,\C^4)$ with the same domain as $\DS$ by the Kato-Rellich Theorem. Moreover, by the resolvent identity
$$
  (\A-i)^{-1} = (\DS - i)^{-1} - (\A-i)^{-1} (\A - \DS) (\DS - i)^{-1}
$$
we infer that the operator $\A$ has also compact resolvent since $\A - \DS$ is bounded. Finally, note that the operator $\A$ commutes with the self-adjoint operators $D_\varphi$ and $D_\psi$ equipped with anti-periodic boundary conditions. There exists thus a sequence of vector-valued functions $Y_{lnm} = U_{lnm}(\theta) e^{in\varphi} e^{im\psi}$ where $l \in \Z^*$ and $n,m \in \frac{1}{2} + \Z$ which forms a Hilbert basis of $L^2(S^3,d\omega;\C^4)$ and are common eigenfunctions of the self-adjoint operators $\A, D_\varphi$ and $D_\psi$. Here the vector-valued functions $U_{lnm}$ are solutions of the eigenvalue equation
\begin{equation} \label{EigenvaluesEq}
  \A^{nm} U_{lnm}(\theta) = \lambda_{lnm}(\omega) U_{lnm}(\theta),
\end{equation}
where the self-adjoint operators $\A^{nm}$ are equal to $\A$ with $D_\varphi$ and $D_\psi$ replaced by $n$ and $m$ in $\frac{1}{2} + \Z$ respectively. 

Note now that the operator $\Go$ (and also $\Ga$) anticommutes with $\A^{nm}$. Therefore, if $Y_{lnm}$ is an eigenfunction of $\A^{nm}$ associated to $\lambda_{lnm}(\omega)$, then $\Go Y_{lnm}$ is an eigenfunction of $\A^{nm}$ associated to $-\lambda_{lnm}(\omega)$. The spectrum of $\A^{nm}$ is thus symmetric with respect to $0$ and we can adopt the following convention. For all $l \in \N^*$, the eigenfunctions $Y_{lnm}$ correspond to the \emph{positive} eigenvalues $\lambda_{lnm}(\omega)$ listed in increasing order, \textit{i.e.} $\lambda_{1nm} \leq \lambda_{2nm} \leq \dots$. Note in passing that for $(n,m) \in \frac{1}{2} + \Z$ fixed, the eigenvalues $\lambda_{lnm}$ of $\A^{nm}$ are \emph{simple} since the Dirac type operator $\A^{nm}$ is limit point at $\theta = 0$ and $\theta = \frac{\pi}{2}$. Moreover, for all $l \in -\N^*$, we set $\lambda_{lnm} = - \lambda_{(-l)nm}$ and $Y_{lnm} = \Go Y_{(-l)nm}$.  Furthermore, the operator $\gamma^1 = -i \Go \Ga$ commutes with $\A$ since $\Go$ and $\Ga$ anticommutes with $\A$. Denoting by $P^\pm = \frac{1}{2} (I_4 \pm \gamma^1)$ the projectors onto the positive and negative spectrum of $\gamma^1$ respectively, we see that the eigenfunctions $Y_{lnm}$ belong either to $Ran P^+$, or $Ran P^-$. We can thus classify the eigenfunctions $Y_{lnm}$ into $Y_{lnm}^+ = P^+ Y_{lnm}$ and $Y_{lnm}^- = P^- Y_{lnm}$. 

Summarising the previous remarks, we have the following decomposition of $L^2(S^3,d\omega;\C^4)$.
\begin{eqnarray*}
  L^2(S^3,d\omega;\C^4) & = & \bigoplus_{l \in \Z^*,(n,m) \in \frac{1}{2} + \Z} \langle Y_{lnm} \rangle, \\ 
                        & = & \bigoplus_{l \in \N^*, (n,m) \in \frac{1}{2} + \Z} \langle Y_{lnm}, \Go Y_{lnm} \rangle, \\ 
                        & = & \bigoplus_{l \in \N^*, (n,m) \in \frac{1}{2} + \Z} \langle P^+ Y_{lnm}, P^- Y_{lnm}, P^+ \Go Y_{lnm}, P^- \Go Y_{lnm} \rangle.  
\end{eqnarray*}
Hence, the Hilbert space $\H$ can be written as
\begin{eqnarray*}
  \H & = & L^2(\R,dx) \otimes L^2(S^3,d\omega;\C^4), \\
     & = & \bigoplus_{l,n,m \in \mathcal{L}} \Big( L^2(\R,dx) \otimes \langle P^+ Y_{lnm}, P^- Y_{lnm}, P^+ \Go Y_{lnm}, P^- \Go Y_{lnm} \rangle \Big), \\
     & := & \bigoplus_{l,n,m \in \mathcal{L}} \H_{lnm},  
\end{eqnarray*}
where $\mathcal{L} = \{ (l,n,m), \ l \in \N^*, \ n,m \in \frac{1}{2} + \Z \}$. By definition of the subspaces $\H_{lnm}$, it is clear that each element $U_{lnm} \in \H_{lnm}$ can be written as
$$
  U_{lnm} = u_{lnm}^1(x) P^+ Y_{lnm} + u_{lnm}^2(x) P^- Y_{lnm} + u_{lnm}^3(x) P^+ \Go Y_{lnm} + u_{lnm}^4(x) P^- \Go Y_{lnm},
$$
with scalar functions $u_{lnm}^j$. Therefore, identifying $U_{lnm}$ with the vector-valued function $^t(u_{lnm}^1,u_{lnm}^2,u_{lnm}^3,u_{lnm}^4)$, we see that the $\H_{lnm}$'s are isometric to $L^2(\R,dx;\C^4)$.  

With this identification at hand and using that
$$
  \Ga P^\pm = i P^\mp \Go, \quad \gamma^1 P^\pm = \pm P^\pm, \quad \Go P^\pm = P^\mp \Go,
$$
it is an easy calculation to see that the $\H_{lnm}$'s are let invariant through the actions of $\A, D_\varphi, D_\psi, \Ra$ and that, for all $U_{lnm} = \ ^t(u_{lnm}^1,u_{lnm}^2,u_{lnm}^3,u_{lnm}^4)$, 
\begin{eqnarray*}
  \A U_{lnm} & = & \lambda_{lnm}(\omega) M_2 U_{lnm}, \quad D_\varphi U_{lnm} = n U_{lnm}, \quad D_\psi U_{lnm} = m U_{lnm}, \\
  \Ra U_{lnm} & = & \Big[ \frac{(r^2+a^2)(r^2+b^2)}{r\sqrt{\Delta}} M_1 D_x + mr M_0 + \Big( \frac{an(r^2+b^2)}{r\sqrt{\Delta}} + \frac{an}{r} M_0 \Big)  + \Big( \frac{bm(r^2+a^2)}{r\sqrt{\Delta}} + \frac{bm}{r} M_0 \Big) \\
              &   & \hspace{2cm} + \frac{ab}{r^2} M_3 - \omega \Big( \frac{(r^2+a^2)(r^2+b^2)}{r\sqrt{\Delta}} + \frac{ab}{r} M_0 \Big) \Big] U_{lnm}, 
\end{eqnarray*}
where the $M_j$ are $4 \times 4$-matrices given by
$$
  M_1 = \left( \begin{array}{cc} 0&\sigma_2\\ \sigma_2&0 \end{array} \right), \quad M_2 = \left( \begin{array}{cc} I_2&0\\0&-I_2 \end{array} \right), \quad M_3 = \left( \begin{array}{cc} \sigma_3\\ 0&\sigma_3 \end{array} \right), \quad M_4 = \left( \begin{array}{cc} 0&\sigma_1 \\ \sigma_1&0 \end{array} \right), 
$$
and the $\sigma_j$'s are the usual Pauli matrices given by
$$
  \sigma_1 = \left( \begin{array}{cc} 0&1 \\ 1&0 \end{array} \right), \quad \sigma_2 = \left( \begin{array}{cc} 0&-i \\ i&0 \end{array} \right), \quad \sigma_3 = \left( \begin{array}{cc} 1&0 \\ 0&-1 \end{array} \right).
$$
Now, choosing the $\gamma^j$'s Dirac matrices in such a way that $\Ga = M_1$, $\Gb = M_2$, $\Go = M_4$ (which would imply that $\gamma^1 = -i\Go \Ga = M_3$), then the Lemma is proved. If we choose any other representation of the $\gamma^j$'s Dirac matrices satisfying (\ref{Clifford}), then there would exist a nonsingular $4 \times 4$ matrix $S$ such that
$$
  S \Ga S^{-1} = M_1, \quad S \Gb S^{-1} = M_2, \quad S \gamma^1 S^{-1} = M_3, \quad S \Go S^{-1} = M_4,
$$ 
(see \cite{Th}). Thus, any representation of the Dirac matrices are equivalent and the precise decomposition of the Lemma still holds if we consider the subspaces $\tilde{H}_{lnm} = S \H_{lnm}$ instead of $\H_{lnm}$. Without loss of generality, we can thus always assume that $\Ga = M_1$, $\Gb = M_2$, $\Go = M_4$ and $\gamma^1 = -i\Go \Ga = M_3$. 

As a last remark, notice that the same proof applies to Lemma \ref{SphericalHarmonics} directly. It suffices to replace the operator $\A$ by the simpler operator $\DS$ in the previous proof. In that case, the positive eigenvalues $\lambda_{lnm}$ of $\DS$ are explicitely computable and are known to be given by $\{\frac{3}{2}, \frac{5}{2}, \frac{7}{2}, \dots \}$ (see for instance \cite{CH}). \\ $\diamondsuit$ \\

{\bf Acknowledgements:} This research was supported by NSERC Discovery Grant 105490-2004 and a James McGill Grant.


\end{document}